\documentclass[a4paper,12pt]{article}

\usepackage[margin=2.5cm]{geometry}
\usepackage{graphicx}
\usepackage{amssymb}
\usepackage{multirow}
\usepackage{amsmath}
\usepackage{setspace}
\usepackage{caption}
\usepackage{subcaption}
\usepackage{float}
\usepackage{mathrsfs}
\usepackage[nosort]{cite}

\newcommand{\dif}{\mathrm{d}}

\newcommand{\Eqref}[1]{(\ref{#1})}
\newcommand{\half}{\frac{1}{2}}

\newcommand{\brac}[1]{\left(#1 \right)}
\newcommand{\sbrac}[1]{\left[#1\right]}

\newcommand{\eq}{\,=\,}

\makeatletter
\renewcommand\section{\@startsection {section}{1}{\z@}%
                                   {-3.5ex \@plus -1ex \@minus -.2ex}%
                                   {2.3ex \@plus.2ex}%
                                   {\normalfont\bfseries}}
\renewcommand\subsection{\@startsection{subsection}{2}{\z@}%
                                     {-3.25ex\@plus -1ex \@minus -.2ex}%
                                     {1.5ex \@plus .2ex}%
                                     {\normalfont\it}}% from \large
\renewcommand\subsubsection{\@startsection{subsubsection}{3}{\z@}%
                                     {-3.25ex\@plus -1ex \@minus -.2ex}%
                                     {1.5ex \@plus .2ex}%
                                     {\normalfont}}% from \normalsize
\makeatother

\begin{document}

\thispagestyle{empty}
\begin{flushright}

\end{flushright}
\vbox{}
\vspace{2cm}

\begin{center}
{\LARGE{A new form of the C-metric with cosmological constant%\\[2mm]
        }}\\[16mm]
{{Yu Chen,~~Yen-Kheng Lim~~and~~Edward Teo}}
\\[6mm]
{\it Department of Physics,
National University of Singapore, %\\[1mm]
Singapore 119260}\\[15mm]

\end{center}
\vspace{2cm}

\centerline{\bf Abstract}
\bigskip
\noindent
The new form of the C-metric proposed by Hong and Teo, in which the two structure functions are factorised, has proved useful in its analysis. In this paper, we extend this form to the case when a cosmological constant is present. The new form of this solution has two structure functions which are partially factorised; moreover, the roots of the structure functions are now regarded as fundamental parameters. This leads to a natural representation of the solution in terms of its so-called domain structure, in which the allowed coordinate range can be visualised as a ``box'' in a two-dimensional plot. The solution is then completely parameterised by the locations of the edges of this box, at least in the uncharged case. We also briefly analyse other possible domain structures---in the shape of a triangle and trapezoid---that might describe physically interesting space-times within the AdS C-metric.

\newpage

\section{Introduction} \label{intro}

The C-metric is a static solution to the vacuum Einstein field equations, whose history dates back to 1918 when it was discovered by Levi-Civita \cite{Levi-Civita}. It was subsequently rediscovered by various other authors in the early 1960's \cite{Newman,Robinson,Ehlers}; in particular, it was Ehlers and Kundt \cite{Ehlers} who, in the process of classifying degenerate static vacuum solutions, gave it the `C' designation that it is known by today. However, its interpretation remained obscure until 1970, when Kinnersley and Walker \cite{Kinnersley:1970zw} showed that the C-metric actually describes a Schwarzschild black hole undergoing uniform acceleration. It was also these two authors who introduced the well-known form of the C-metric that would remain the {\it de facto\/} standard form for the next three decades or so.

To see how Kinnersley and Walker obtained their form of the C-metric, we need to start with the slightly more general form used by Ehlers and Kundt \cite{Ehlers}:
\begin{align}
\dif s^2=\frac{1}{(x-y)^2}\left[F(y)\,\dif t^2-\frac{\dif y^2}
{F(y)}+\frac{\dif x^2}{G(x)}+G(x)\,\dif\phi^2\right],\label{ehlers_form}
\end{align}
where the structure functions $G(x)$ and $F(y)$ are cubic polynomials in $x$ and $y$ respectively, satisfying the condition
\begin{align}
F(x)=G(x)\,.\label{F=G}
\end{align}
Thus the two polynomials share the same coefficients.
It would appear that this solution has four parameters, which can be taken to be the coefficients of $G(x)$, say. However, two of them are actually unphysical, and can be gauged away by a suitable coordinate transformation. Kinnersley and Walker considered the following affine coordinate transformation:
\begin{align}
x'=Ac_0x+c_1\,,\quad y'=Ac_0y+c_1\,,\quad t'=c_0t\,,\quad \phi'=c_0\phi\,,
\label{coord_freedom}
\end{align}
under which the metric (\ref{ehlers_form}) gains an overall factor but otherwise retains the same general form:
\begin{align}
\dif s^2=\frac{1}{A^2(x-y)^2}\left[F(y)\,\dif t^2-\frac{\dif y^2}
{F(y)}+\frac{\dif x^2}{G(x)}+G(x)\,\dif\phi^2\right].\label{KW_metric_form}
\end{align}
Note that the structure functions $G(x)$ and $F(y)$ are still cubic polynomials satisfying (\ref{F=G}), although with new coefficients depending on $A$, $c_0$ and $c_1$. Kinnersley and Walker then used the coordinate freedom in (\ref{coord_freedom}) to set $G(x)$ to be
\begin{align}
G(x)=1-x^2-2mAx^3.\label{KW_form}
\end{align}
In particular, the linear coefficient has been set to zero. The parameters $m$ and $A$ are related to the mass and acceleration of the black hole respectively. In the limit $A\rightarrow0$, the usual Schwarzschild metric with mass parameter $m$ can be recovered from this form of the C-metric. On the other hand, in the limit $m\rightarrow0$, the usual Rindler space metric with acceleration parameter $A$ can be recovered.
 
A major disadvantage of the Kinnersley--Walker form of the C-metric is that the roots of the structure function (\ref{KW_form}) are cumbersome to write down in terms of the parameters $m$ and $A$. Nevertheless, knowledge of these roots is important, since they encode the locations of the axes and horizons in the space-time. Almost any study of the geometrical properties of the space-time will involve these roots, and would be very complicated as a result. Even if the roots were not explicitly expressed in terms of $m$ and $A$, one would need to have a handle on their dependence on these parameters.

In 2003, Hong and Teo \cite{Hong:2003gx} proposed a new form of the C-metric that would alleviate this difficulty. Instead of using the coordinate freedom in (\ref{coord_freedom}) to set the linear coefficient of $G(x)$ to zero, they used this freedom to set it to the value $2mA$. As a result, $G(x)$ can be put in the factorised form:
\begin{align}
G(x)=(1-x^2)(1+2mAx)\,.\label{new_form}
\end{align}
In this form, the roots of the structure functions are obvious to read off: the two axes of the space-time are located at $x=\pm1$, while the acceleration and black-hole horizons are located at $y=-1$, $-\frac{1}{2mA}$, respectively. These simple expressions lead to potentially drastic simplifications when analysing the properties of the C-metric, as demonstrated in \cite{Hong:2003gx}.

The new form (\ref{new_form}) is related to the previous one (\ref{KW_form}) by a coordinate transformation and redefinition of parameters. In particular, $m$ and $A$ still retain their interpretations as the mass and acceleration parameters of the black hole respectively. Again, the Schwarzschild metric can be recovered in the limit $A\rightarrow0$, while the Rindler space metric can be recovered in the limit $m\rightarrow0$. However, we emphasise that in the general case $m,A\neq0$, the parameters appearing in (\ref{new_form}) are inequivalent to those appearing in (\ref{KW_form}).

The C-metric can be straightforwardly extended to include charge, by adding a quartic term to the structure functions. In the Kinnersley--Walker form, the metric is still given by (\ref{KW_metric_form}), but the structure function (\ref{KW_form}) is generalised to
\begin{align}
G(x)=1-x^2-2mAx^3-q^2A^2x^4,\label{KW_form2}
\end{align}
where $q$ is the charge parameter of the black hole. Being a quartic polynomial, the roots of $G(x)$ are now even more cumbersome to write down than in the vacuum case. Fortunately, the factorised form (\ref{new_form}) can be extended to the charged case. It was shown in \cite{Hong:2003gx} that, by a coordinate transformation and redefinition of parameters, (\ref{KW_form2}) can be written as
\begin{align}
G(x)=(1-x^2)(1+r_+Ax)(1+r_-Ax)\,,\label{new_charged_form}
\end{align}
where $r_\pm=m\pm\sqrt{m^2-q^2}$ are the locations of the horizons in the usual form of the Reissner--Nordstr\"om metric. In this form, the roots of $G(x)$ are trivial to read off: the two axes of the space-time are again located at $x=\pm1$, while the acceleration and two black-hole horizons are located at $y=-1$, $-\frac{1}{r_\pm A}$, respectively.

The (charged) C-metric can also be extended to include rotation. In this case, the metric (\ref{KW_metric_form}) has to be replaced by a more complicated stationary form---not reproduced here---which nevertheless still depends on two structure functions $G(x)$ and $F(y)$ satisfying (\ref{F=G}). In the Kinnersley--Walker form, $G(x)$ is given by
\begin{align}
G(x)=1-x^2-2mAx^3-(a^2+q^2)A^2x^4,\label{KW_form3}
\end{align}
where $a$ is the rotation parameter of the black hole. In \cite{Hong:2004dm}, Hong and Teo showed that $G(x)$ can again be written in the factorised form (\ref{new_charged_form}), but with $r_\pm=m\pm\sqrt{m^2-a^2-q^2}$. The latter are just the locations of the horizons in the Boyer--Lindquist form of the Kerr--Newman metric. However, as Hong and Teo pointed out, one key difference in this case is that this new form of the rotating C-metric is {\it not\/} related to the traditional form (\ref{KW_form3}) by a coordinate transformation. It turns out that the traditional form of the rotating C-metric possesses so-called Dirac--Misner singularities along the axes, while the new form does not. To avoid such singularities, the structure functions necessarily take the factorised form (\ref{new_charged_form}).

A natural question at this stage is whether this new form of the (static, charged) C-metric can be extended to include a cosmological constant $\Lambda$. The C-metric with cosmological constant is traditionally written in the form (\ref{KW_metric_form}), with the structure functions
\begin{align}
  G(x)&=1-{x}^2-2 m A{x}^3-q^2A^2x^4,\nonumber\\
  F(y)&=\brac{1-\frac{1}{\ell^2A^2}}-{y}^2-2 m A{y}^3-q^2A^2y^4,\label{CC_SF}
\end{align}
where $\ell^2\equiv-{3}/{\Lambda}$.
Note that $G(x)$ has exactly the same form as in (\ref{KW_form2}), but that $F(x)$ now differs from $G(x)$ by a constant term:
 \begin{align}
F(x)=G(x)-\frac{1}{\ell^2A^2}\,.
\end{align}
This implies that there is no simple relation between the roots of $G(x)$ and those of $F(y)$. In particular, a factorised form for $G(x)$ does not lead to one for $F(y)$, or {\it vice versa\/}. In \cite{Hong:2004dm}, a tentative proposal was made to write $G(x)$ in the factorised form (\ref{new_charged_form}), at the expense of leaving $F(y)$ unfactorised. However, an unsatisfactory consequence is that the $r_\pm$ appearing in $G(x)$ have no relation to the locations of the horizons of the Kerr--Newman--dS/AdS black hole. This is perhaps not unexpected, since the locations of the horizons are encoded by the roots of $F(y)$, which as mentioned are now not the same as those of $G(x)$.

In this paper, we would like to find a new form of the C-metric with cosmological constant that retains the nice features of the factorised form of \cite{Hong:2003gx}. To this end, recall that two of the roots of $G(x)$ are physically significant, in that they represent the two axes in the space-time. The coordinate range for $x$ lies between these two roots. On the other hand, two of the roots of $F(y)$ are physically significant, in that they represent the acceleration and (outer) black-hole horizons. The coordinate range for $y$ lies between these two roots. It is therefore natural to take these two roots of $G(x)$ and two roots of $F(y)$ {\it as parameters of the solution\/}. This would lead to a partial factorisation of $G(x)$ and of $F(y)$, in the sense that they are of the form:
\begin{align}
  G(x)&=(x-\alpha)(x-\beta)(\;\cdots)\,,\nonumber\\
  F(y)&=(y-a)(y-b)(\;\cdots)\,,
\end{align}
where the ellipses denote quadratic (in the uncharged case, linear) polynomials in either $x$ or $y$. This form of the structure functions resolves the dilemma of whether one should completely factorise $G(x)$ at the expense of leaving $F(y)$ unfactorised, or {\it vice versa\/}.

For simplicity, let us focus on the uncharged case. When expressed in terms of the parameters $\alpha$, $\beta$, $a$ and $b$, we will see that the metric can be written in the form (\ref{Cmetric_alphabeta}) below. Note in particular that the cosmological-constant parameter $\ell^2$ appears as a conformal factor of the metric. This is in contrast to the traditional form (\ref{KW_metric_form}), in which it is the acceleration parameter $A^{-2}$ that appears as a conformal factor. We also remark that the four roots $\alpha$, $\beta$, $a$ and $b$ are analogous to the four coefficients of $G(x)$ parameterising the Ehlers--Kundt form of the Ricci-flat C-metric (\ref{ehlers_form}). The key difference is that we are now parameterising the polynomials in terms of their {\it roots\/} rather than their coefficients.

At this stage, we can use a coordinate transformation of the form (\ref{coord_freedom}) (with $A=1$) to set $\alpha=-1$ and $\beta=+1$, say. The resulting metric, whose explicit form can be found in (\ref{Cmetric}) below, is then parameterised in terms of the remaining roots $a$ and $b$, as well as the cosmological-constant parameter $\ell$. The two parameters $a$ and $b$ play the role of $m$ and $A$ in the traditional form of the C-metric given by (\ref{KW_metric_form}) and (\ref{CC_SF}) with $q=0$. It is important to note that this represents a shift in paradigm. Instead of trying to parameterise the C-metric in terms of properties of the black hole ($m$ and $A$), we now regard the roots of the structure functions ($a$ and $b$) as fundamental parameters. The properties of the black hole, if needed, can then be expressed in terms of these roots.

The new form of the C-metric with cosmological constant (\ref{Cmetric}) may seem somewhat longer than the traditional form given by (\ref{KW_metric_form}) and (\ref{CC_SF}) with $q=0$, but we shall see that the analysis of this solution will be simpler and more transparent in the new form. For example, its parameterisation in terms of the four roots $\alpha$, $\beta$, $a$ and $b$ leads to a natural way to visualise the allowed ranges of the coordinates $x$ and $y$ in an $x$-$y$ plot. Since $x$ ranges between $\alpha$ and $\beta$, and $y$ ranges between $a$ and $b$, they will fill out a rectangle or ``box'' in the $x$-$y$ plot.\footnote{Such plots are, of course, not new. They have appeared, for instance, in \cite{Pravda:2000vh} for the Ricci-flat case, \cite{Podolsky:2000pp} for the case of positive cosmological constant, and \cite{Podolsky:2002nk} for the case of negative cosmological constant. What is new here is the unified treatment of all three cases, by treating the edges of the box as parameters of the solution. We will also make use of such plots to classify the different space-times described by the C-metric.} We will refer to it as the domain of the space-time.\footnote{This is not to be confused with the domain structure of a higher-dimensional black-hole space-time introduced in \cite{Harmark:2009dh}.} By the above-mentioned coordinate freedom, the two sides of the box can be fixed to be at $x=\pm1$. However, the upper and lower edges of the box are free to vary, and their positions will parameterise the solution. We will see in particular that the location of the upper edge determines the sign of the cosmological constant. Examples of the domains in the case of a positive, negative and zero cosmological constant can be found in Fig.~\ref{NRBfig2} below, in the darker shade.

We have already mentioned that the two sides of the box at $x=\pm1$ represent the two axes of the space-time, while the upper and lower edges represent the acceleration and black-hole horizons respectively. The domain of the space-time may also be bounded by the line $x=y$, which represents asymptotic infinity, as in the case of negative cosmological constant in Fig.~\ref{NRBfig2}(b). Indeed, the structure of the domains in the various cases already gives much useful geometrical information about the space-time without the need for detailed calculations.  

One of the major benefits of studying the domain structure is that it allows for a complete classification of the possible space-times described by the C-metric. Although in this paper, we are primarily interested in space-times with two axes and two horizons, it is clear that other possibilities are allowed. For example, in Fig.~\ref{NRBfig2}(b), the triangular-shaped corner of the box that is cut off by the $x=y$ line describes a space-time with just one axis and one horizon. The existence of such new space-times is difficult to infer from the traditional form of the C-metric, and they are almost impossible to classify in any case.

We mention that the addition of charge does not change much of what has been described so far. The solution, whose explicit form can be found in (\ref{Cmetric_charged}) and (\ref{Cmetric_charged_max}) below, is now parameterised by $a$, $b$ and $q$, in addition to $\ell$. In particular, the domain of the space-time can still be visualised as a box in an $x$-$y$ plot, with $a$ and $b$ the locations of its lower and upper edges respectively. This box is in fact exactly the same as the one in the uncharged case, since the addition of charge does not affect the four roots $\alpha$, $\beta$, $a$ and $b$.

This paper is organised as follows: We begin in Sec.~\ref{sec2} by deriving the new form of the uncharged C-metric with cosmological constant. In Sec.~\ref{sec3}, we analyse the space-time geometry described by this general solution, in particular, the axes and horizons contained in it. To this end, we will study its domain structure, as well as its so-called rod structure, in detail. In Sec.~\ref{sec4}, we specialise to the three cases of a positive, zero and negative cosmological constant. In each case, we obtain the full parameter range of the solution and describe how various known limits can be obtained. The charged generalisation of the C-metric with cosmological constant is discussed in Sec.~\ref{sec5}. In Sec.~\ref{sec6}, we discuss other possible domains that can emerge from the C-metric with a negative cosmological constant. The paper then concludes with a brief discussion of future work. There is also an appendix detailing the coordinate transformation from the new form of the solution to the traditional form (\ref{KW_metric_form}) and (\ref{CC_SF}), at least for the uncharged case.

\section{Derivation of the new form}
\label{sec2}

We begin with the following form of the C-metric with a cosmological constant, taken as a static limit of the Pleba\'nski--Demia\'nski metric \cite{Plebanski:1976gy}\footnote{This is obtained from Eq.~(6.3) of \cite{Plebanski:1976gy} by redefining $p=x$, $q=-y$, $\sigma=\phi$, $\tau=t$, $\epsilon'=\epsilon$, $n'=n$, $m'=m$, $e'=e$ and $g'=g$, $-\Lambda/6+\gamma'= \gamma_1$ and $\Lambda/6+\gamma'= \gamma_2$. Additionally, the structure functions are redefined as $\mathscr{P}(p)=\mathcal{P}(x)$ and $\mathscr{Q}(q)=-\mathcal{Q}(y)$.}:
\begin{align}
 \dif s^2&\eq\frac{1}{(x-y)^2}\brac{\mathcal{Q}(y)\,\dif t^2-\frac{\dif y^2}{\mathcal{Q}(y)}+\frac{\dif x^2}{\mathcal{P}(x)}+\mathcal{P}(x)\,\dif\phi^2},\nonumber\\
 \mathcal{P}(x)&\eq\gamma_1+2nx-\epsilon x^2+2mx^3-q^2x^4,\nonumber\\
 \mathcal{Q}(y)&\eq\gamma_2+2ny-\epsilon y^2+2my^3-q^2y^4, \label{static_PD_metric}
\end{align}
where $q^2\equiv e^2+g^2$, with $e$ and $g$ being the electric and magnetic charges respectively. The corresponding Maxwell potential is
\begin{align}
 \mathcal{A}\eq ey\,\dif t-gx\,\dif\phi\,. \label{static_PD_max}
\end{align}
Together, (\ref{static_PD_metric}) and (\ref{static_PD_max}) are a solution to the Einstein--Maxwell equations with cosmological constant $\Lambda=3(\gamma_2-\gamma_1)$. It will turn out to be convenient to define
\begin{align}
\Lambda\equiv-\frac{3}{\ell^2}\,,
\end{align}
where $\ell^2$ can take either sign. The so-called de Sitter (dS) case in which $\Lambda>0$ then corresponds to $\ell^2<0$, while the anti-de Sitter (AdS) case in which $\Lambda<0$ corresponds to $\ell^2>0$. The Ricci-flat case in which $\Lambda=0$ is recovered in the limit $\ell^2\rightarrow\pm\infty$.

We first consider the uncharged case $e=g=0$. In this case, the structure functions $\mathcal{P}(x)$ and $\mathcal{Q}(y)$ become cubic polynomials. The parameters of the solution can then be taken to be the four polynomial coefficients of $\mathcal{P}(x)$, in addition to the cosmological constant. As described in the introduction, we wish to reparameterise the solution in terms of two roots of $\mathcal{P}(x)$ and two roots of $\mathcal{Q}(y)$. Thus we write the structure functions in the form
\begin{align}
 \mathcal{P}(x)&=(x-\alpha)(x-\beta)(w_0+w_1x)\,,\nonumber\\
 \mathcal{Q}(y)&=(y-a)(y-b)(k_0+k_1y)\,.\label{factor_ansatz}
\end{align}
By comparing the polynomial coefficients of \Eqref{factor_ansatz} with \Eqref{static_PD_metric}, we can express $w_0$, $w_1$, $k_0$ and $k_1$ in terms of $\alpha$, $\beta$, $a$ and $b$. If we further pull out a constant conformal factor in the metric and rescale the coordinates $t$ and $\phi$ appropriately, the result is
\begin{align}
  \dif s^2&\eq\frac{-\ell^2(a-\alpha)(a-\beta)(b-\alpha)(b-\beta)}{(x-y)^2}\brac{Q(y)\,\dif t^2-\frac{\dif y^2}{Q(y)}+\frac{\dif x^2}{P(x)}+P(x)\,\dif\phi^2},\nonumber\\
  P(x)&\eq(x-\alpha)(x-\beta)\sbrac{(a+b-\alpha-\beta)(x-a-b)+ab-\alpha\beta},\nonumber\\
  Q(y)&\eq(y-a)(y-b)\sbrac{(a+b-\alpha-\beta)(y-\alpha-\beta)+ab-\alpha\beta}.\label{Cmetric_alphabeta}
\end{align}
Although this form of the metric is longer than the one in (\ref{static_PD_metric}), it has the key advantage that all the roots of the structure functions can be explicitly read off:
\begin{align}
 P(x)=0:&\quad\alpha,\;\beta,\;\mbox{and }\;\gamma\eq a+b+\frac{\alpha\beta-ab}{a+b-\alpha-\beta}\,,\nonumber\\
 Q(y)=0:&\quad a,\;b,\;\;\mbox{and }\;c\eq\alpha+\beta+\frac{\alpha\beta-ab}{a+b-\alpha-\beta}\,.
\label{roots}
\end{align}
As desired, the metric is now parameterised by the four roots $\alpha$, $\beta$, $a$ and $b$, in addition to $\ell$. In this paper, we shall restrict ourselves to the case in which each of the structure functions admits at least two real roots. This in fact implies that the third root of each structure function is also real by (\ref{roots}). We remark that it is possible for $\alpha$ and $a$ to be complex; a real metric is still obtained provided $\beta=\alpha^*$ and $b=a^*$. In this case, $P(x)$ and $Q(y)$ will have exactly one real root each, given by $\gamma$ and $c$ respectively.

At this stage, we note that the metric \Eqref{Cmetric_alphabeta} has the following two continuous symmetries and one discrete symmetry:
\begin{itemize}
 \item Translational symmetry:

The metric is invariant under
  \begin{align}
   x&\rightarrow x+c_0\,,\quad y\rightarrow y+c_0\,,\quad\alpha\rightarrow \alpha+c_0\,,\nonumber\\
   \beta&\rightarrow \beta+c_0\,,\quad a\rightarrow a+c_0\,,\quad b\rightarrow b+c_0\,, \label{trans}
  \end{align}
  for any choice of constant $c_0$. This corresponds to a translation of the $x$ and $y$ coordinates.

 \item Rescaling symmetry:

The metric is invariant under
  \begin{align}
   x&\rightarrow c_1x\,,\quad y\rightarrow c_1y\,,\quad t\rightarrow\frac{t}{c_1^3}\,,\quad\phi\rightarrow\frac{\phi}{c_1^3}\,,\nonumber\\
   \alpha&\rightarrow c_1\alpha\,,\quad\beta\rightarrow c_1\beta\,,\quad a\rightarrow c_1a\,,\quad b\rightarrow c_1b\,, \label{scale}
  \end{align}
  for any choice of constant $c_1\neq 0$. In particular, if $c_1<0$, this corresponds to a reflection---in addition to a rescaling---of the $x$ and $y$ coordinates.

 \item Parameter symmetry:

The metric is invariant under the interchange of any pair of roots of $P(x)$:
  \begin{align}
    \alpha\leftrightarrow\beta\,,\quad\alpha\leftrightarrow\gamma\,,\quad\mbox{or}\quad\beta\leftrightarrow\gamma\,;
  \end{align}
and similarly under the interchange of any pair of roots of $Q(y)$:
  \begin{align}
    a\leftrightarrow b\,,\quad a\leftrightarrow c\,,\quad\mbox{or}\quad b\leftrightarrow c\,.
  \end{align}
While the metric is obviously invariant under $\alpha\leftrightarrow\beta$ or $a\leftrightarrow b$, it is not so for interchanges involving $\gamma$ or $c$. To show, for example, the invariance of the metric under $\alpha\leftrightarrow\gamma$, one has to solve for $\alpha$ in terms of $\gamma$ and substitute it into (\ref{Cmetric_alphabeta}). One can then show that it is brought back to exactly the same form (\ref{Cmetric_alphabeta}) after the relabelling $\gamma\rightarrow\alpha$ and a rescaling of the coordinates $t$ and $\phi$. 

\end{itemize}

For completeness, we also note that, if Wick-rotations are allowed, \Eqref{Cmetric_alphabeta} has an additional discrete symmetry:
\begin{itemize}
  \item Coordinate symmetry:

The metric is invariant under the interchanges
  \begin{align}
   x\leftrightarrow y\,,\quad t\leftrightarrow \phi\,,\quad\alpha\leftrightarrow a\,,\quad\beta\leftrightarrow b\,,\label{coord}
  \end{align}
  followed by a double-Wick rotation $t\rightarrow it$ and $\phi\rightarrow i\phi$.
\end{itemize}

Now, we would eventually be interested in regions of space-time with the correct Lorentzian signature. To ensure that the signature does not change in the region of interest, note that $x$ must lie between a pair of adjacent roots of $P(x)$, and similarly $y$ must lie between a pair of adjacent roots of $Q(y)$.\footnote{The ranges of $x$ and $y$ are necessarily finite, because as we shall see, there are curvature singularities at $x=\pm\infty$ and $y=\pm\infty$.}  By parameter symmetry, we can set these roots to be $\alpha$ and $\beta$, and $a$ and $b$ respectively, so that
\begin{align}
\alpha< x< \beta\,,\quad a< y< b\,.\label{box_region0}
\end{align}
Furthermore, we can use the two continuous symmetries above to set two of the four roots to specific values. A natural choice is to set $\alpha=-1$ and $\beta=+1$, which we shall adopt from now on. Thus the region of interest is now
\begin{align}
-1< x<+1\,,\quad a< y< b\,. \label{box_region}
\end{align}

With this choice of parameters, the resulting metric takes the slightly simpler form:
\begin{align}
 \dif s^2&\eq\frac{H^2}{(x-y)^2}\brac{Q(y)\,\dif t^2-\frac{\dif y^2}{Q(y)}+\frac{\dif x^2}{P(x)}+P(x)\,\dif\phi^2},\nonumber\\
 P(x)&\eq\brac{x^2-1}\sbrac{(a+b)(x-a-b)+ab+1},\nonumber\\
 Q(y)&\eq(y-a)(y-b)\sbrac{(a+b)y+ab+1}, \label{Cmetric}
\end{align}
where we have introduced the abbreviation\footnote{As we shall see, the constant $H^2$ is always positive in the cases we are interested in.}
\begin{align}
 H^2\equiv\ell^2\brac{a^2-1}\brac{1-b^2}.\label{H2}
\end{align}
There are now just three independent parameters: $a$, $b$ and $\ell$. This is the new form of the C-metric with cosmological constant that will be studied in detail below. In this form, the roots of the structure functions are
\begin{align}
 P(x)=0:&\quad \pm 1,\;\mbox{and }\;\gamma\eq\frac{a^2+b^2+ab-1}{a+b}\,,\nonumber\\
 Q(y)=0:&\quad a,\;b,\;\mbox{and }\; c\eq-\frac{ab+1}{a+b}\,. \label{NRB_roots}
\end{align}

Finally, we note that (\ref{Cmetric}) is invariant under the reflection
\begin{align}
 x\rightarrow -x\,,\quad y\rightarrow -y\,,\quad a\rightarrow-a\,,\quad b\rightarrow -b\,.
\end{align}
This is actually a residual symmetry from (\ref{scale}) that is present even after fixing the values of $\alpha$ and $\beta$. It can be used to set 
\begin{align}
 a+b< 0\,, \label{aplusb}
\end{align}
without loss of generality.\footnote{It can be shown that if $a+b=0$, the region of interest (\ref{box_region}) leads to a space-time with the wrong signature. So this special case can be ruled out.}
Since we can always take $a<b$, it follows from (\ref{aplusb}) that $a<0$.

\section{Domain and rod-structure analysis}
\label{sec3}

We now turn to an analysis of the space-time geometry described by (\ref{Cmetric}), in the range (\ref{box_region}).

We first require that the space-time has the correct Lorentzian signature. Recall that the range of $x$ lies between two adjacent roots of $P(x)$, while that of $y$ lies between two adjacent roots of $Q(y)$. In this range of coordinates, $P(x)$ must have opposite sign to $Q(y)$ to ensure a Lorentzian signature. These conditions can be used to fix the ordering of the roots to one of two possibilities, as we shall now show.

\begin{figure}[t]
  \begin{center}
   \begin{subfigure}[b]{0.4\textwidth}
    \centering
    \includegraphics[scale=0.8]{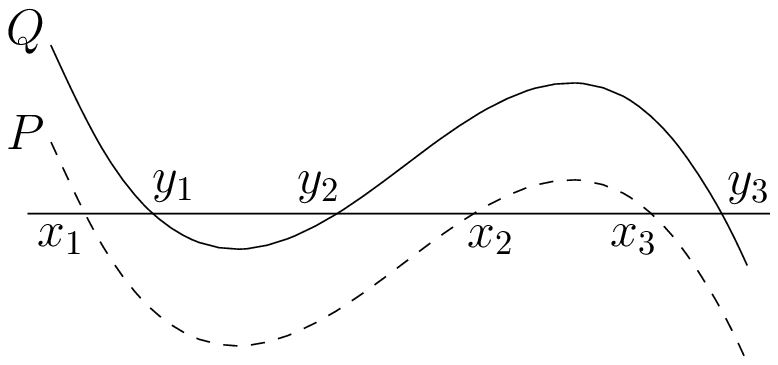}
    \caption{$\brac{a^2-1}\brac{1-b^2}<0$}
   \end{subfigure}
~~~   \begin{subfigure}[b]{0.4\textwidth}
    \centering
    \includegraphics[scale=0.8]{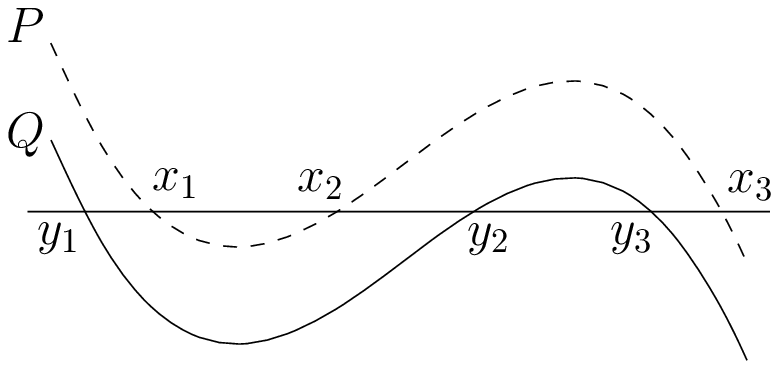}
    \caption{$\brac{a^2-1}\brac{1-b^2}>0$}
   \end{subfigure}
  \end{center}
  \caption{The curves of $P(x)$ and $Q(x)$ for (a) $P(x)-Q(x)<0$, and (b) $P(x)-Q(x)>0$. The solid lines represent $Q$, while dashed lines represent $P$.}
  \label{NRBfig1}
 \end{figure}

Note that the structure functions $P(x)$ and $Q(y)$ obey the identity
\begin{align}
 P(x)-Q(x)\eq&\brac{a^2-1}\brac{1-b^2}. \label{PQ_diff}
\end{align}
In other words, they share the same polynomial coefficients except for the constant term which differs by the right-hand side of (\ref{PQ_diff}). In particular, their leading coefficient is negative by (\ref{aplusb}). The form of $P(x)$ and $Q(x)$ is sketched in Fig.~\ref{NRBfig1}, separately for the cases $\brac{a^2-1}\brac{1-b^2}\lessgtr 0$.

In both Fig.~\ref{NRBfig1}(a) and (b), we have denoted the roots of $P(x)$ in increasing order by $\{x_1,x_2,x_3\}$, and those of $Q(y)$ also in increasing order by $\{y_1,y_2,y_3\}$. A specific ordering of these six roots is implied in either case. We would now like to identify the roots $\{x_1,x_2,x_3\}$ with the possible choices $\{-1,+1,\gamma\}$, and similarly identify $\{y_1,y_2,y_3\}$ with the possible choices $\{a,b,c\}$. Bearing in mind the ranges in (\ref{box_region}) as well as the condition (\ref{aplusb}), we then have the following two possibilities:
\begin{itemize}
 \item For the case $\brac{a^2-1}\brac{1-b^2}<0$, the only possibility is to identify $(x_1=\gamma,\break x_2=-1,x_3=+1)$ and $(y_1=a,y_2=b,y_3=c)$. These six roots are then ordered as
  \begin{align}
   \gamma< a < b <-1<+1< c\,. \label{NRBorder_b}
  \end{align}
In this case $P(x)>0$ and $Q(y)<0$ in the region of interest (\ref{box_region}), and the correct $(-+++)$ signature is obtained only if $\ell^2<0$. This corresponds to a positive cosmological constant, i.e., the dS case.
 \item For the case $\brac{a^2-1}\brac{1-b^2}>0$, the only possibility is to again identify $(x_1=\gamma,\break x_2=-1,x_3=+1)$ and $(y_1=a,y_2=b,y_3=c)$. These six roots are then ordered as
  \begin{align}
   a<\gamma< -1<b< c<+1\,. \label{NRBorder_a}
  \end{align}
In this case again $P(x)>0$ and $Q(y)<0$ in the region of interest (\ref{box_region}), and the correct $(-+++)$ signature is obtained only if $\ell^2>0$. This corresponds to a negative cosmological constant, i.e., the AdS case.
\end{itemize}
Note that in either case, we have $H^2>0$. We also note that the order of the roots of $P(x)$ and those of $Q(y)$ are the same in either case, i.e., $\gamma<-1<+1$ and $a<b<c$. It is the relative order of these two sets of roots that are different in the two cases. 

Since $\gamma$ and $c$ depend on $a$ and $b$ as in (\ref{NRB_roots}), the ordering in (\ref{NRBorder_b}) or (\ref{NRBorder_a}) will in general imply an even more restricted range for $a$ and $b$. Since this will depend on which case we are considering, the details will be left to the next section where we treat the dS and AdS cases separately.

At this stage, we recall that the Ricci-flat case in which the cosmological constant vanishes can be obtained from either of the above two cases by taking the limit $\ell^2\rightarrow\pm\infty$. In this case, the six roots are ordered as
\begin{align}
 \gamma= a < b =-1<+1= c\,. \label{NRBorder_c}
\end{align}
The positivity of $H^2$ still holds in this case, although it is to be understood as a limit in which $\ell^2\rightarrow\pm\infty$ and $b\rightarrow-1$ from above or below.

\begin{figure}
 \begin{center}
  \begin{subfigure}[b]{0.4\textwidth}
   \centering
   \includegraphics[height=2.4in,width=2.4in]{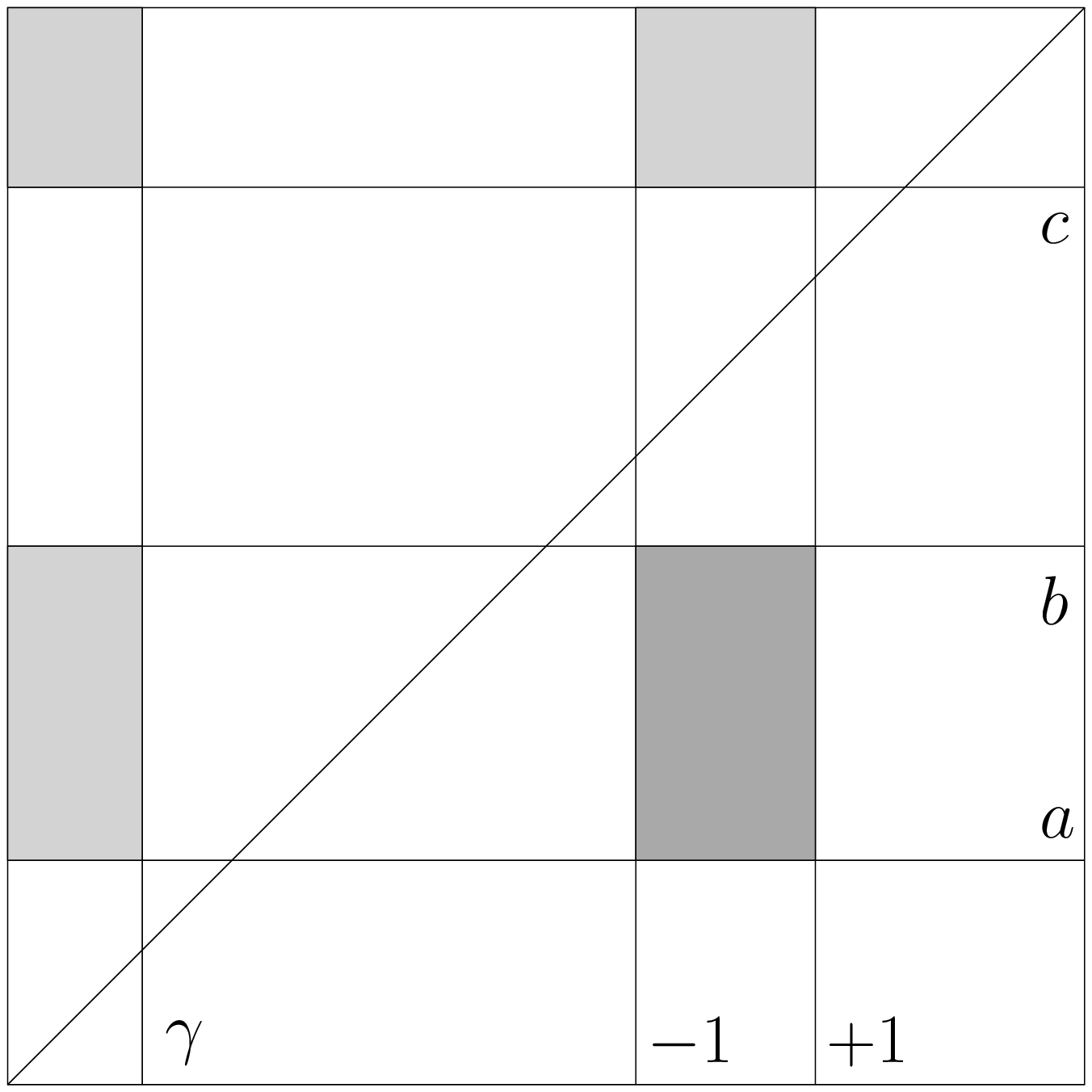}
   \caption{$\ell^2<0$ (dS)}
   \end{subfigure}\vspace{6pt}
~~~  \begin{subfigure}[b]{0.4\textwidth}
   \centering
   \includegraphics[height=2.4in,width=2.4in]{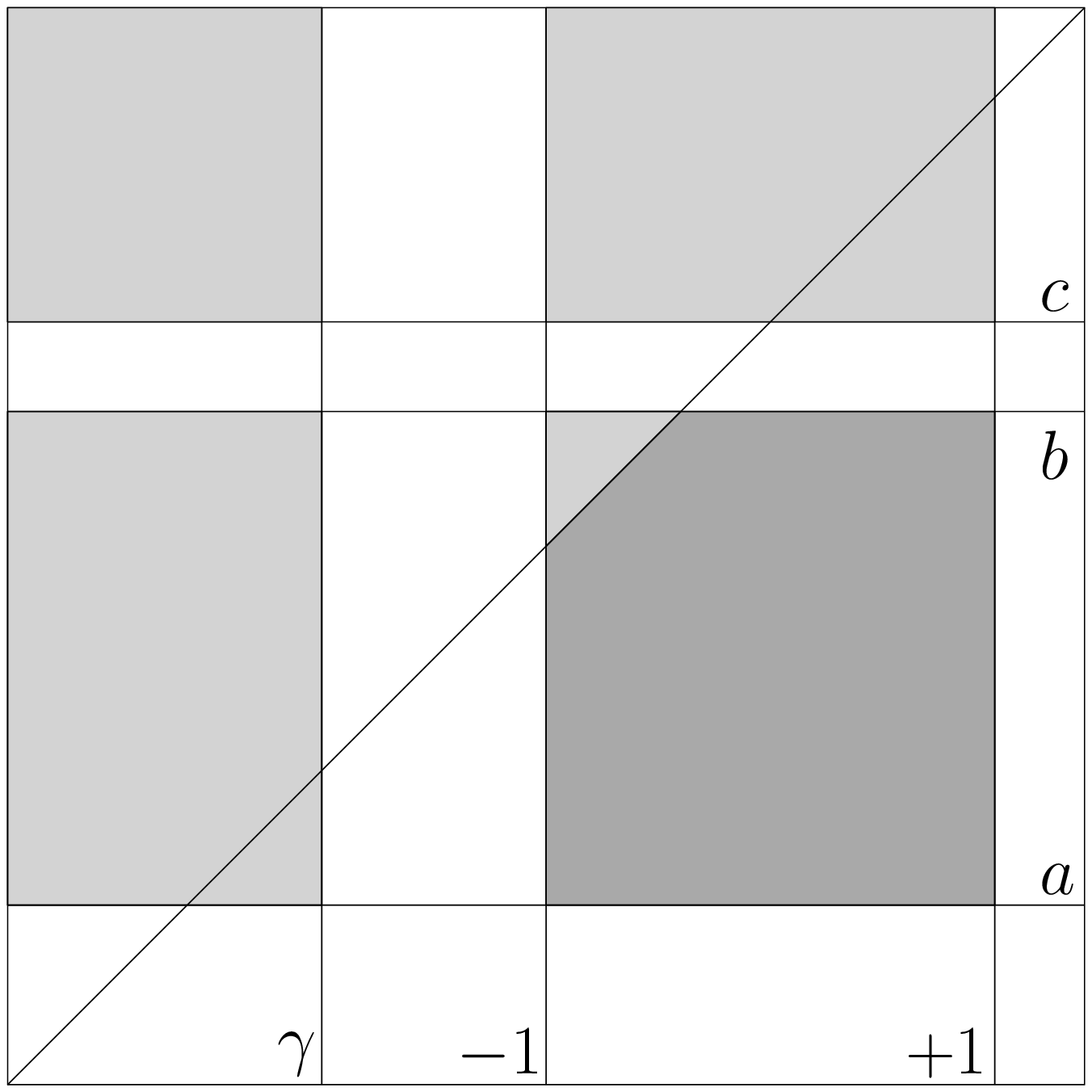}
   \caption{$\ell^2>0$ (AdS)}
   \end{subfigure}\par\medskip
  \begin{subfigure}[b]{0.4\textwidth}
   \centering
   \includegraphics[height=2.4in,width=2.4in]{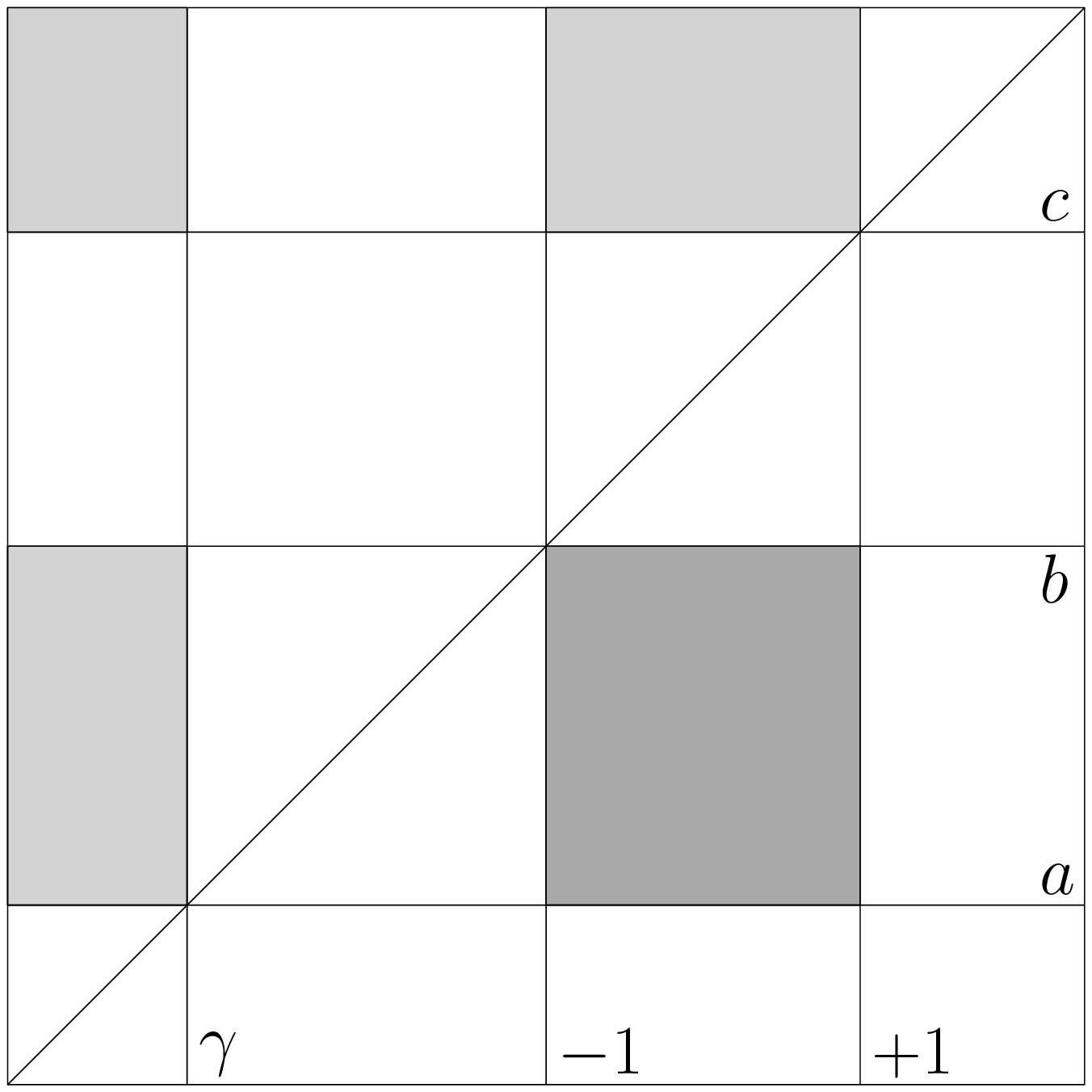}
   \caption{$\ell^2\rightarrow\pm\infty$ (Ricci-flat)}
   \end{subfigure}
 \end{center}
 \caption{The domains for the cases (a) $\ell^2<0$ (dS), (b) $\ell^2>0$ (AdS), and (c) $\ell^2\rightarrow\pm\infty$ (Ricci-flat). In each diagram, the $x$-axis is in the horizontal direction, while the $y$-axis is in the vertical direction. The boxes we are interested in are in the darker shade. Regions in a lighter shade also have the correct signature, but are not the main focus here. The diagonal lines correspond to asymptotic infinity.}
 \label{NRBfig2}
\end{figure}

Having determined the order of the roots in each of the three cases (dS, AdS and Ricci-flat), we are now in a position to plot these roots in a standard $x$-$y$ plot. The roots of $P(x)$ will be vertical lines in this plot, while the roots of $Q(y)$ will be horizontal lines. These two sets of lines will intersect each other, dividing the $x$-$y$ space into rectangles or ``boxes''. The range of interest (\ref{box_region}) will then be one of these boxes, which we refer to as the domain of the space-time. The domain for each of the three cases is illustrated in Fig.~\ref{NRBfig2} in the darker shade. Also indicated in the plots in a lighter shade are other regions with the correct Lorentzian signature, although we will not be focussing on them in this paper.

The physical interpretation of the boundaries of the domain will be discussed below when we study the rod structure of the space-time. We only mention here that in general roots of $P(x)$ correspond to symmetry axes in the space-time, while roots of $Q(y)$ correspond to Killing horizons in the space-time. The space-time may also end at asymptotic infinity or at a curvature singularity, if they are present. It is clear from \Eqref{Cmetric} that asymptotic infinity is located at $x=y$. This corresponds to the diagonal line in each of the plots in Fig.~\ref{NRBfig2}.

To check the presence of curvature singularities, we turn to the Kretschmann invariant of the metric \Eqref{Cmetric}:
\begin{align}
 R_{\mu\nu\lambda\sigma}R^{\mu\nu\lambda\sigma}\eq\frac{24}{\ell^4}+\frac{12(a+b)^2(x-y)^6}{\ell^4\brac{a^2-1}^2\brac{1-b^2}^2}\,. \label{NRB_Kret}
\end{align}
We see that there are curvature singularities located at $x,y\rightarrow\pm\infty$. Only in special cases (such as $a\rightarrow -\infty$) is the Kretschmann invariant everywhere finite and the metric \Eqref{Cmetric} becomes trivially (A)dS space.

Thus there are curvature singularities at the edges of each of the plots in Fig.~\ref{NRBfig2}, if the edges are taken to represent $x,y=\pm\infty$. Any region touching these edges will then have curvature singularities. We see that by restricting to the finite range (\ref{box_region}), our space-time will be guaranteed to be free of curvature singularities. On the other hand, it is possible for the domain to intersect the $x=y$ line, as can be seen from the plots in Fig.~\ref{NRBfig2}. In the dS case in which $b<-1$, the box will not touch this line at all. But in the Ricci-flat case in which $b=-1$, one corner of the box will touch this line. In the AdS case in which $b>-1$, the box will necessarily intersect the line. The domain in the latter case is then given by (\ref{box_region}) with the added restriction $y<x$. We will still refer to it below as a ``box'', albeit one with a corner cut off. It is continuously connected to the boxes in the Ricci-flat and dS cases.

It should be noted that the triangular-shaped corner that is cut off by $x=y$ in Fig.~\ref{NRBfig2}(b) represents a new space-time, distinct from the one represented by the rest of the box. Like the latter, it has the correct Lorentzian signature and is free of curvature singularities, and so might have an interesting physical interpretation. We will return to this triangular domain in Sec.~\ref{sec6}, when we classify all the possible shapes that can be taken by a domain with the correct signature and without curvature singularities.

We now turn to a study of the four edges of the box, at which $x=\pm1$ and $y=a,b$. Note that these are points at which the metric coefficient $g_{\phi\phi}$ or $g_{tt}$ vanishes, i.e., at which the norm of the Killing vector field $\frac{\partial}{\partial\phi}$ or $\frac{\partial}{\partial t}$ becomes zero. These sets of points have the interpretation of either the axes or horizons of the space-time. Recently, a powerful formalism, known as the rod-structure formalism, has been developed to study such points in a unified way. In this formalism, the sets of points form a series of ``rods'' each with a certain defined direction, and the rod structure of a space-time will contain much useful information about its physical and geometrical properties. We would like to investigate the rod structure of the metric (\ref{Cmetric}). 

The rod-structure formalism was originally defined for four-dimensional Ricci-flat space-times with two commuting Killing vector fields. The reader is referred to \cite{Chen:2010zu}, and references therein, for more details of this formalism. When a cosmological constant is present, most of the features of this formalism will remain the same as in the Ricci-flat case. In particular, the definitions of a turning point and the normalised direction of a rod are still applicable. However, one difference arising from the presence of a cosmological constant is that the canonical Weyl--Papapetrou coordinates can no longer be used to describe a rod structure. It has to be described using some other coordinates; for the metric \Eqref{Cmetric}, the $(x,y)$ coordinates can be used. 

\begin{figure}[!t]

\begin{center}
\includegraphics{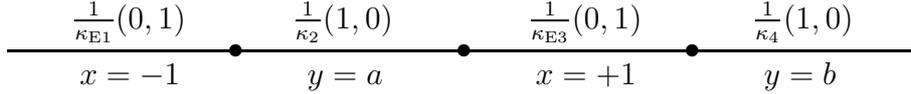}
\caption{The rod structure of the metric \Eqref{Cmetric}. The location of each rod is indicated below it, while its direction is indicated above it. The three solid points are turning points where adjacent rods meet. In the dS case, the left and right end-points of the rod structure should be identified at a fourth turning point. In the Ricci-flat and AdS cases, the left-most and right-most rods are infinite in extent.}
\label{rod_diagram}
\end{center}

\end{figure}

Focusing on the coordinate range (\ref{box_region}), the rod structure of \Eqref{Cmetric} consists of the following four rods:
\begin{itemize}
 \item Rod 1: a space-like rod located at $(x=-1,\; a\leq y\leq b)$, with (normalised) direction\footnote{Here, the direction $(\alpha,\beta)$ is defined by $\alpha\frac{\partial}{\partial t}+\beta\frac{\partial}{\partial\phi}$.} $\ell_1=\frac{1}{\kappa_{\mathrm{E1}}}(0,1)$, where
  \begin{align}
   \kappa_{\mathrm{E1}}=a^2+b^2+ab+a+b-1\,.\label{direction_uncharged1}
  \end{align}
 \item Rod 2: a time-like rod located at $(-1\leq x\leq+1, y=a)$, with direction $\ell_2=\frac{1}{\kappa_2}(1,0)$, where
  \begin{align}
   \kappa_2=\half(b-a)\brac{1+a^2+2ab}.\label{direction_uncharged2}
  \end{align}
 \item Rod 3: a space-like rod located at $(x=+1,\;a\leq y\leq b)$, with direction $\ell_3=\frac{1}{\kappa_{\mathrm{E3}}}(0,1)$, where
  \begin{align}
   \kappa_{\mathrm{E3}}=a^2+b^2+ab-a-b-1\,.\label{direction_uncharged3}
  \end{align}
 \item Rod 4: a time-like rod located at $(-1\leq x\leq+1,\; y=b)$, with direction $\ell_4=\frac{1}{\kappa_4}(1,0)$, where
  \begin{align}
   \kappa_4=\half(b-a)\brac{1+b^2+2ab}.\label{direction_uncharged4}
  \end{align}
\end{itemize}
These four rods meet at the three turning points $(x=-1,y=a)$, $(x=1,y=a)$ and $(x=1,y=b)$. This rod structure is schematically illustrated in Fig.~\ref{rod_diagram}. It can be seen that Rods 1 and 3 correspond to the left and right edges of a box respectively, while Rods 2 and 4 correspond to the lower and upper edges respectively. In a sense, the rod structure can be ``folded up'' by joining the two endpoints together, to form the four edges of the box.\footnote{In the case of the triangular domain mentioned above, as well as the trapezoidal ones to be described in Sec.~\ref{sec6}, the rod structure will be different and can be obtained accordingly. In particular, it will consist of just two or three rods, which can be folded up to form the two or three edges of the corresponding triangular or trapezoidal domain.}

In the dS case, it can be checked that all four rods are finite, in the sense that the axes or horizons they represent are finite in extent. On the other hand, Rods 1 and 4 are infinite in the Ricci-flat case. This is consistent with the box diagram in Fig.~\ref{NRBfig2}(c), whereby the upper-left corner of the box touches the $x=y$ line representing asymptotic infinity. Rods 1 and 4 are also infinite in the AdS case. This is again consistent with the box diagram in Fig.~\ref{NRBfig2}(b), whereby the left and upper edges of the box intersect the $x=y$ line. In this case, Rods 1 and 4 should only consist of the parts $(x=-1,\; a\leq y< -1)$ and $(b< x\leq+1,\; y=b)$ respectively.

Now, in the rod-structure formalism, time-like rods represent Killing horizons while space-like rods represent symmetry axes. For a time-like rod with direction $\frac{1}{\kappa}\frac{\partial}{\partial t}$, the normalisation factor $\kappa$ is nothing but the surface gravity of the corresponding horizon. It also gives the temperature of the horizon via the relation
\begin{align}
T=\frac{\kappa}{2\pi}\,.
\end{align}
On the other hand, for a space-like rod with direction $\frac{1}{\kappa_{\rm E}}\frac{\partial}{\partial \phi}$, the normalisation factor $\kappa_{\rm E}$ can be interpreted as the Euclidean surface gravity of this rod. It actually encodes the natural periodicity of the $\phi$ coordinate around the corresponding axis. To avoid a conical singularity along this axis, the identification
\begin{align}
(t,\phi)\rightarrow \left(t,\phi+\frac{2\pi}{\kappa_{\rm E}}\right),
\end{align}
should be made.

Armed with these facts, we are now ready to interpret the rod structure of the present solution. We begin with the Ricci-flat case, since this is perhaps the most familiar of the three cases. In this case, Rod 2 is a black-hole horizon, while Rod 4 is an acceleration horizon. This is confirmed by the fact that the former is finite in extent, while the latter is infinite in extent. $\kappa_2$ and $\kappa_4$ are their respective surface gravities. On the other hand, Rod 3 is an inner axis that separates the black-hole and acceleration horizons, while Rod 1 is an outer asymptotic axis. This is again confirmed by the fact that the former is finite in extent, while the latter is infinite in extent. $\kappa_{\rm E3}$ and $\kappa_{\rm E1}$ are their respective Euclidean surface gravities. A similar interpretation of these four rods applies to the AdS case.

In the dS case, the main difference is that Rods 1 and 4 are now finite in extent. This is in agreement with the naive expectation that a cosmological horizon will appear in this case and separate the region of interest from asymptotic infinity. It is the cosmological horizon that now plays the role of the acceleration horizon.

From the fact that $a\pm b<0$ in general, we immediately have that
\begin{align}
\kappa_{\rm E1}< \kappa_{\rm E3}\,,\quad \kappa_{\rm 2}> \kappa_{\rm 4}\,.\label{order_surface_gravities_vacuum}
\end{align}
The latter implies that the temperature of the acceleration (or cosmological) horizon is always lower than that of the black-hole horizon. On the other hand, the fact that $\kappa_{\rm E1}\neq \kappa_{\rm E3}$ implies it is not possible to find a global identification of the $\phi$ coordinate that would avoid conical singularities along Rods 1 and 3 simultaneously. Thus the space-time cannot be made completely regular: one has to allow for the presence of a conical singularity either on the inner axis (Rod 3) or on the asymptotic axis (Rod 1). It is this conical singularity which acts as the source of acceleration for the black hole.

Finally, we remark that in this paper we are only interested in static regions of the space-time, which would confine us to the regions in between the black-hole and acceleration (or cosmological) horizons. It is possible to find non-static coordinates extending past say the black-hole horizon, into the region inside the black hole. This region is described by the box that lies immediately below the dark-shaded box in each of the plots in Fig.~\ref{NRBfig2}.  As can be seen, there is necessarily a curvature singularity in this region at $y=-\infty$, corresponding to the black-hole singularity.

\section{Parameter ranges and various limits}
\label{sec4}

While we have completed the analysis of the space-time geometry of the C-metric \Eqref{Cmetric}, there remains a few details to tidy up. One such detail is the full ranges of the parameters $a$ and $b$, which will depend on the specific case (dS, AdS or Ricci-flat) we are considering. Thus we will discuss the three cases separately in this section. With the full parameter range, we will then identify various known limits in each case, and interpret them in terms of the box diagrams in Fig.~\ref{NRBfig2}.

\subsection{dS C-metric}

In this case, we have $\ell^2<0$, and the ordering of the roots of the structure functions is given by (\ref{NRBorder_b}). Since the third roots $\gamma$ and $c$ depend on $a$ and $b$ as in (\ref{NRB_roots}), the full range of $a$ and $b$ will have to be consistent with this ordering. It turns out that the full parameter range of the dS C-metric is simply
\begin{align}
-\infty< a<b<-1\,, \label{dS_range}
\end{align}
which will ensure that $\gamma<a$ and $c>1$. It can be checked that for this range, the surface gravities (\ref{direction_uncharged1})--(\ref{direction_uncharged4}) are all positive.

We now briefly describe what happens at the various limits of the range (\ref{dS_range}). When $a\rightarrow-\infty$, the metric \Eqref{Cmetric} reduces to de Sitter space. To see this, we first need to perform the rescaling
\begin{align}
 x\rightarrow\sqrt{\ell^2(1-b^2)}\,x\,,\quad y\rightarrow\sqrt{\ell^2(1-b^2)}\,y\,,\quad t\rightarrow\frac{t}{a^2\sqrt{\ell^2(1-b^2)}}\,,\quad\phi\rightarrow\frac{\phi}{a^2\sqrt{\ell^2(1-b^2)}}\,,\label{massless_limit1}
\end{align}
before taking the limit $a\rightarrow-\infty$. The metric then becomes
\begin{align}
 \dif s^2&\eq\frac{1}{(x-y)^2}\brac{-F(y)\,\dif t^2+\frac{\dif y^2}{F(y)}+\frac{\dif x^2}{G(x)}+G(x)\,\dif\phi^2},\nonumber\\
     G(x)&=A^2-x^2,\quad F(y)= y^2-A^2+\frac{1}{\ell^2}\,,\quad A^2=\frac{1}{\ell^2(1-b^2)}\,,\label{massless_limit2}
\end{align}
which is nothing but de Sitter space in a disguised form. Note that the cosmological horizon is located at $y=-\sqrt{A^2-\frac{1}{\ell^2}}$.

In terms of the box diagram in Fig.~\ref{NRBfig2}(a), this limit corresponds to sending the lower edge of the box to $-\infty$, while keeping its upper edge fixed. This effectively removes the black hole from the picture. One can calculate the area of the black-hole horizon, and show that it indeed becomes zero in this limit. In terms of the rod structure in Fig.~\ref{rod_diagram}, the second rod shrinks down to zero length and disappears in this limit.

Another possible limit is to set the acceleration of the black hole to zero, in which case we should recover the Schwarzschild-de Sitter black hole. This is achieved by taking the limit $a,b\rightarrow-\infty$, such that their ratio remains constant. To see this, we define $\mu\equiv \frac{b}{a}$, and the new coordinates
\begin{align}
 x=\cos\theta\,,\quad y=\sqrt{\frac{-\ell^2}{1+\mu+\mu^2}}\frac{a\mu}{r}\,,\quad\phi\rightarrow \frac{\phi}{a^2\brac{1+\mu+\mu^2}}\,,\quad t\rightarrow\sqrt{\frac{-\ell^2}{1+\mu+\mu^2}}\frac{t}{a^3\mu\ell^2}\,.
\end{align}
If we take the limit $a\rightarrow -\infty$ while keeping $\mu$ constant, then \Eqref{Cmetric} reduces to the familiar form of the Schwarzschild--de Sitter metric:
\begin{align}
 \dif s^2=-\brac{1-\frac{2m}{r}+\frac{r^2}{\ell^2}}\dif t^2+\brac{1-\frac{2m}{r}+\frac{r^2}{\ell^2}}^{-1}\dif r^2+r^2\brac{\dif\theta^2+\sin^2\theta\,\dif\phi^2},
\end{align}
with
\begin{align}
 m\equiv\frac{\mu(1+\mu)}{2\brac{1+\mu+\mu^2}}\sqrt{\frac{-\ell^2}{1+\mu+\mu^2}}\,.
\end{align}
From the parameter range \Eqref{dS_range}, it follows that $0<\mu< 1$, and thus we have
\begin{align}
 0< m<\sqrt{\frac{-\ell^2}{27}}\,.
\end{align}
Hence from our form, one can recover all possible Schwarzschild--de Sitter black holes with positive mass and with both black-hole and cosmological horizons present.

This limit has an intuitive understanding in terms of the box diagram in Fig.~\ref{NRBfig2}(a). It corresponds to sending both the upper and lower edges of the box to $-\infty$, which is the only way in which the left and right edges of the box can become symmetric. Physically, this situation is needed to ensure that the space-time regions around the north and south poles of the black hole become symmetric in this limit. 

Finally, at the other endpoint of the range (\ref{dS_range}), we have the limit $b\rightarrow-1$. We have already noted that this gives the C-metric with no cosmological constant. This limit will be examined next.

\subsection{Ricci-flat C-metric}
\label{sec4.2}

In this case, we have $\ell^2\rightarrow\pm\infty$, and the ordering of the roots of the structure functions is given by (\ref{NRBorder_c}). Note that this limit requires $b$ to take the value $-1$, and this in turn fixes the values of the third roots to be $\gamma=a$ and $c=1$. The remaining free parameter $a$ then takes the full range
\begin{align}
-\infty<a<-1\,,\label{ricci_flat_range}
\end{align}
without any further restriction.

Since the general form \Eqref{Cmetric} is not valid in this limit, we need to seek an alternative form of the C-metric in the Ricci-flat case. We first reparameterise $b$ as
\begin{align}
 b\eq-\sqrt{1+\frac{k^2}{\ell^2(1+a)}}\,,
\end{align}
where $k^2$ is a positive constant. Then we have $H^2=(1-a)k^2$, which is manifestly positive when $a$ takes the range (\ref{ricci_flat_range}). Upon taking the limit $\ell^2\rightarrow\pm\infty$ and after an appropriate rescaling of the coordinates $t$ and $\phi$, we obtain
\begin{align}
 \dif s^2&\eq\frac{k^2}{(x-y)^2}\brac{G(y)\,\dif t^2-\frac{\dif y^2}{G(y)}+\frac{\dif x^2}{G(x)}+G(x)\,\dif\phi^2},\cr
G(x)&\eq\brac{1-x^2}\brac{x-a}.\label{old_C}
\end{align}
This form is in fact identical to the factorised form of the Ricci-flat C-metric that was first proposed in \cite{Hong:2003gx}. An exact correspondence to the form used in that paper can be made if the parameters are redefined as
\begin{align}
 k^2=\frac{1}{2mA^3}\,,\quad a=-\frac{1}{2mA}\,,\label{corresp}
\end{align}
and a further rescaling of $t$ and $\phi$ is performed.

As the various limits of the Ricci-flat C-metric are well understood in this factorised form \cite{Hong:2003gx,Griffiths:2006tk,Griffiths:2009dfa}, we shall not discuss them here.

\subsection{AdS C-metric}

In this case, we have $\ell^2>0$, and the ordering of the roots of the structure functions is given by (\ref{NRBorder_a}). For this ordering to be consistent with the expressions of the third roots in \Eqref{NRB_roots}, $a$ and $b$ have to take a more restricted parameter range. It turns out that the allowed parameter range is the union of
\begin{align}
{\Bigg\{}\begin{array}{ll}
-1<b\leq -a-\sqrt{a^2-1}\,, &a\leq -\frac{5}{3}\,;\\
-1<b\leq -\frac{1+a+\sqrt{5-2a-3a^2}}{2}\,,&-\frac{5}{3}\leq a<-1\,.
  \end{array}\label{AdS_range1}
\end{align}
It can be checked that for this range, the surface gravities (\ref{direction_uncharged1})--(\ref{direction_uncharged4}) are all non-negative.

\begin{figure}
 \begin{center}
  \includegraphics[scale=0.9]{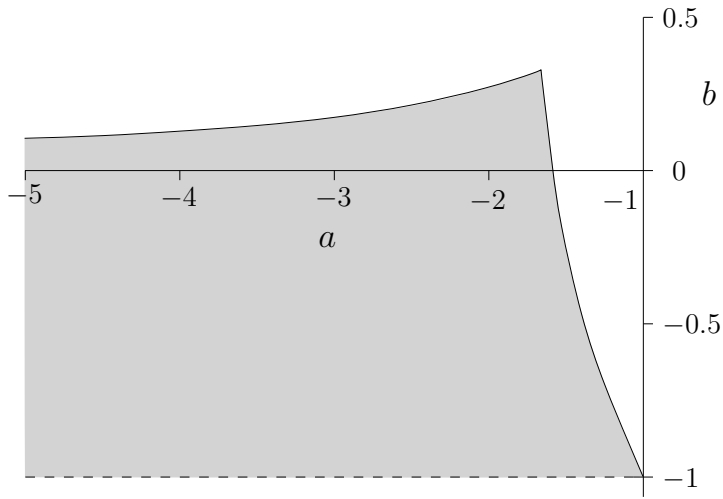}
 \end{center}
 \caption{The parameter range of the AdS C-metric, as a plot of $b$ versus $a$. The left boundary of this range extends to $a=-\infty$. The intersection point of the upper and right boundary curves is at $(a=-\frac{5}{3},b=\frac{1}{3})$.}
 \label{NRBfig3}
\end{figure}
\par

It is illuminating to visualise the parameter range (\ref{AdS_range1}) in an $a$-$b$ plot, as in Fig.~\ref{NRBfig3}. As can be seen, the boundary of this parameter range consists of four separate parts. We now briefly describe what happens as each of these different boundaries is approached.

The lower boundary $b\rightarrow-1$ corresponds to the Ricci-flat limit, which was discussed in the previous subsection. 

The left boundary $a\rightarrow-\infty$ corresponds to the limit in which the AdS C-metric reduces to anti-de Sitter space. This limit is taken in the same way as in the dS case, i.e., we first perform the rescaling \Eqref{massless_limit1}, and then take the $a\rightarrow-\infty$ limit to obtain \Eqref{massless_limit2}. When $\ell^2>0$, it can be shown that the space-time described by \Eqref{massless_limit2} is just anti-de Sitter space, in an accelerating coordinate system whose origin is undergoing a constant acceleration $A$.

However, unlike the dS case, there is no limit of (\ref{Cmetric}) in which we can recover the Schwarzschild--AdS black hole. Since the range of $b$ is bounded in \Eqref{AdS_range1}, one can no longer take the limit $\left|b\right|\rightarrow\infty$. Hence our parameter range of the AdS C-metric does not include the Schwarzschild--AdS black hole as a special case.
This result is perhaps not surprising, since in our analysis, we presume the existence of both the black-hole and acceleration horizons when we require the domain to be a box with four sides. On the other hand, the usual structure function of the Schwarzschild--AdS metric is $1-2m/r+r^2/\ell^2$, which has only one real root (horizon) when $\ell^2>0$. Hence it is not possible to recover the Schwarzschild--AdS black hole from the range of parameters we have identified.\footnote{The Schwarzschild--AdS black hole can be recovered from (\ref{Cmetric}), if $b$ is allowed to take complex values, and if $c=b^*$. In Fig.~\ref{NRBfig2}(b), this corresponds to letting the two roots $y=b,c$ come together and disappear from the diagram, so that the domain of interest becomes a trapezoid. This case will be discussed briefly in Sec.~\ref{sec6}.}

The remaining two boundaries correspond to the two different upper bounds for $b$ in (\ref{AdS_range1}). When $a\leq -\frac{5}{3}$, this upper bound actually corresponds to the limit in which $b=c$: the second and third roots of $Q(y)$ become degenerate. In this case, $\kappa_4=0$ and the acceleration horizon becomes extremal with zero temperature. This solution has been used, for example, by Emparan et al.\ \cite{Emparan:1999wa} to construct a brane-world black hole. 

On the other hand, when $-\frac{5}{3}\leq a<-1$, the upper bound for $b$ corresponds to the limit in which $\gamma=-1$: the first and second roots of $P(x)$ become degenerate. In this case, $\kappa_{\rm E1}=0$ and the outer axis (Rod 1) becomes extremal in the sense that it becomes infinitely far away from other points in the space-time. This axis can be regarded as a new spatial infinity of the space-time. The solution was then interpreted by Hubeny et al.\ \cite{Hubeny:2009ru} as a ``black droplet'', with two disconnected horizons: one extending to asymptotic infinity and the other extending to the new spatial infinity.

These two curves are in fact related by the coordinate symmetry (\ref{coord}), meaning that they can be mapped to each other by a double-Wick rotation. The intersection point $(a=-\frac{5}{3},b=\frac{1}{3})$ of the two boundary curves can then be interpreted as an ``extremal black droplet'', referring to the fact that the horizon that extends to asymptotic infinity is now extremal.

On the boundary curve with $-\frac{5}{3}\leq a<-1$, the solution is characterised by two para\-meters $a$ and $\ell$. If one performs the following coordinate transformation and parameter redefinition:
\begin{align}
x\rightarrow \frac{3+4\sqrt{3}\,x}{9}\,,\quad y\rightarrow \frac{3+4\sqrt{3}\,y}{9}\,,\quad \lambda=-\frac{(3a+5)(1-3a)^2}{32}\,,
\end{align}
and rescales $t$ and $\phi$ appropriately, the metric is brought to the black-droplet metric studied in \cite{Hubeny:2009ru}:
\begin{align}
\label{metric_HMR}
\dif s^2&=\frac{\ell^2(1+\lambda)}{(x-y)^2}\brac{-{F(y)}\,\dif t^2+\frac{\dif y^2}{{F(y)}}+\frac{\dif x^2}{{G(x)}}+{G(x)}\,\dif \phi^2},\nonumber\\
G(x)&=1-x^2-2\mu x^3,\quad F(y)=\lambda+y^2+2\mu y^3,
\end{align}
with $\mu=\frac{1}{3\sqrt{3}}$. In particular, note that the number of independent parameters matches. The equivalence of the other boundary curve with $a\leq -\frac{5}{3}$ to the form of the AdS C-metric used in \cite{Emparan:1999wa} to construct a brane-world black hole, is left to the reader.

\section{Charged generalisation}
\label{sec5}

\subsection{Derivation of the new form}

We now consider the charged C-metric with cosmological constant, i.e., \Eqref{static_PD_metric} with non-vanishing $e$ and $g$. To obtain a factorised form of the structure functions, we write them as
\begin{align}
 \mathcal{P}(x)&\eq(x-\alpha)(x-\beta)\brac{w_0+w_1x-q^2x^2},\nonumber\\
 \mathcal{Q}(y)&\eq(y-a)(y-b)\brac{k_0+k_1y-q^2y^2}. \label{charged_PD_PQanz}
\end{align}
By comparing the polynomial coefficients of \Eqref{charged_PD_PQanz} with \Eqref{static_PD_metric}, we can express $w_0$, $w_1$, $k_0$ and $k_1$ in terms of $\alpha$, $\beta$, $a$ and $b$. If we further pull out a constant conformal factor in the metric and rescale $t$ and $\phi$ appropriately, the result is
\begin{align}
 \dif s^2&\eq\frac{-\ell^2(a-\alpha)(a-\beta)(b-\alpha)(b-\beta)}{(x-y)^2}\brac{Q(y)\,\dif t^2-\frac{\dif y^2}{Q(y)}+\frac{\dif x^2}{P(x)}+P(x)\,\dif\phi^2},\nonumber\\
 P(x)&\eq(x-\alpha)(x-\beta)\bigl[(a+b-\alpha-\beta)(x-a-b)+ab-\alpha\beta  \nonumber\\
                &\hspace{2cm}+\ell^2(a-\alpha)(a-\beta)(b-\alpha)(b-\beta)q^2(x-a)(x-b)\bigr]\,,\nonumber\\
 Q(y)&\eq(y-a)(y-b)\bigl[(a+b-\alpha-\beta)(y-\alpha-\beta)+ab-\alpha\beta\nonumber\\
                &\hspace{2cm}+\ell^2(a-\alpha)(a-\beta)(b-\alpha)(b-\beta)q^2(y-\alpha)(y-\beta)\bigr]\,, \label{Cmetric_charged_alphabeta}
\end{align}
where recall $q^2\equiv e^2+g^2$. The corresponding Maxwell potential is
\begin{align}
  \mathcal{A}\eq -\ell^2(a-\alpha)(a-\beta)(b-\alpha)(b-\beta)\brac{ey\,\dif t-gx\,\dif\phi}. \label{Cmetric_charged_alphabeta_max}
\end{align}
As desired, the metric is now parameterised by the four roots $\alpha$, $\beta$, $a$ and $b$, in addition to the charge parameters $e$ and $g$ and the cosmological-constant parameter $\ell$.

The above form of the charged C-metric with cosmological constant can be seen to respect two continuous and two discrete symmetries, similar to the uncharged case. It respects the translational symmetry in the form (\ref{trans}) up to a gauge transformation of the Maxwell potential, and the rescaling symmetry (\ref{scale}) provided it is accompanied by the substitutions $e\rightarrow e/c_1^2$, $g\rightarrow g/c_1^2$. The solution also respects the parameter symmetry, i.e., it is invariant under the interchange of any pair of roots of $P(x)$, and similarly under the interchange of any pair of roots of $Q(y)$. Lastly, the coordinate symmetry (\ref{coord}) is respected, provided the double-Wick rotation is accompanied by the substitutions $e\rightarrow ie$, $g\rightarrow ig$.

Now we assume that each of the structure functions admits at least two real roots, as in the uncharged case. We take these real roots to be $\alpha$, $\beta$, $a$ and $b$, and let them define the ranges of $x$ and $y$ as in (\ref{box_region0}). We then use the two continuous symmetries to fix $\alpha=-1$ and $\beta=1$, giving the ranges (\ref{box_region}). After the redefinitions\footnote{\label{footnote_H}These redefinitions are consistent, since $H^2$ as defined in (\ref{H2}) is always positive in the cases we are interested in. We have seen that this is true in the uncharged case, and it continues to remain true when charge is added, since $H^2$ does not depend on $q$. Without loss of generality, we set $H=|H|$ here.} $e\rightarrow e/H$ and $g\rightarrow g/H$, the metric becomes
\begin{align}
 \dif s^2&\eq\frac{H^2}{(x-y)^2}\brac{Q(y)\,\dif t^2-\frac{\dif y^2}{Q(y)}+\frac{\dif x^2}{P(x)}+P(x)\,\dif\phi^2},\nonumber\\
 P(x)&\eq\brac{x^2-1}\bigl[(a+b)(x-a-b)+ab+1-q^2(x-a)(x-b)\bigr]\,,\nonumber\\
 Q(y)&\eq(y-a)(y-b)\sbrac{(a+b)y+ab+1-q^2(y^2-1)}. \label{Cmetric_charged}
\end{align}
The Maxwell potential is accordingly
\begin{align}
  \mathcal{A}\eq H\brac{ey\,\dif t-gx\,\dif\phi}. \label{Cmetric_charged_max}
\end{align}
There are now just five independent parameters: $a$, $b$, $e$, $g$ and $\ell$. In this factorised form, the roots of the structure functions are given by
\begin{align}
 P(x)=0:&\quad x=\pm 1,\;\;\mbox{and }\;x_\pm,\nonumber\\
 Q(y)=0:&\quad y=a,\;b,\;\mbox{and }\;y_\pm,\label{charged_roots}
\end{align}
where
\begin{align}
 x_\pm&\equiv\frac{1}{2q^2}\Bigl\{(1+q^2)(a+b)\pm\sqrt{q^4(a-b)^2-2q^2(a^2+b^2-2)+(a+b)^2}\Bigr\}\,,\nonumber\\
 y_\pm&\equiv\frac{1}{2q^2}\Bigl\{(a+b)\pm\sqrt{4q^4+4q^2(1+ab)+(a+b)^2}\Bigr\}\,.
\end{align}

A key feature of the above parameterisation of the charged C-metric is that the first two roots of $P(x)$ and $Q(y)$ in (\ref{charged_roots}) are exactly the same as in the uncharged case. In particular, this will mean that the locations of the axes and horizons remain unchanged even when charge is present. Only the third and fourth roots, $x_\pm$ and $y_\pm$, depend on the charge parameter $q$.

As in the uncharged case, we shall assume $a+b<0$ without loss of generality. In the limit $q\rightarrow0$, one finds that
\begin{align}
 x_+\rightarrow \gamma\,,\quad y_+\rightarrow c\,,
\end{align}
as expected from \Eqref{NRB_roots}. At the same time, the roots $x_{-}$ and $y_{-}$ will recede to $-\infty$, and the structure functions will become cubic polynomials. It can be shown that $y_{\pm}$ are always real. On the other hand, while $x_{\pm}$ are real in the AdS and Ricci-flat cases, they can be complex in the dS case if $q^2$ is large enough. Henceforth, we shall assume for definiteness that $x_{\pm}$ are real, but will indicate the differences when they become complex.

\subsection{Domain and rod-structure analysis}

As in the uncharged case, we first find the possible orderings of the roots that will ensure that the space-time has the correct Lorentzian signature. To this end, we note that the structure functions still obey the identity (\ref{PQ_diff}), with the two separate cases $\brac{a^2-1}\brac{1-b^2}\lessgtr0$ to consider. At the same time, recall (c.f.\ Footnote \ref{footnote_H}) that $H^2$ is always positive even when charge is present; so the above two cases correspond to $\ell^2\lessgtr0$, respectively.

 \begin{figure}[t]
  \begin{center}
   \begin{subfigure}[b]{0.4\textwidth}
    \centering
    \includegraphics[scale=0.8]{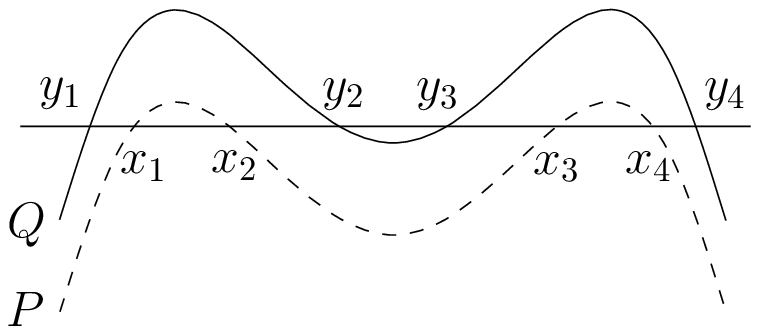}
    \caption{$\brac{a^2-1}\brac{1-b^2}<0$, $\ell^2<0$}
   \end{subfigure}
~~~   \begin{subfigure}[b]{0.4\textwidth}
    \centering
    \includegraphics[scale=0.8]{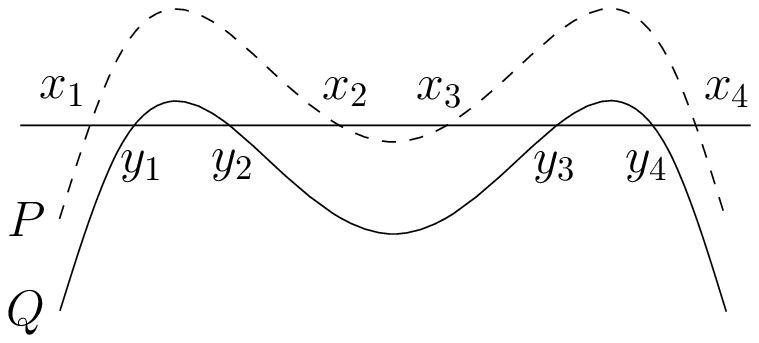}
    \caption{$\brac{a^2-1}\brac{1-b^2}>0$, $\ell^2>0$}
   \end{subfigure}
  \end{center}
  \caption{The curves of $P(x)$ and $Q(x)$ for (a) $P(x)-Q(x)<0$, $\ell^2<0$, and (b) $P(x)-Q(x)>0$, $\ell^2>0$. The solid lines represent $Q$, while dashed lines represent $P$.}
  \label{quartic}
 \end{figure}

These two cases are plotted in Fig.~\ref{quartic}(a) and (b) respectively. In both graphs, we have denoted the roots of $P(x)$ in increasing order by $\{x_1,x_2,x_3,x_4\}$, and those of $Q(y)$ also in increasing order by $\{y_1,y_2,y_3,y_4\}$. Note that Fig.~\ref{quartic}(a) should reduce to Fig.~\ref{NRBfig1}(a) in the uncharged limit, while Fig.~\ref{quartic}(b) should reduce to Fig.~\ref{NRBfig1}(b) in this limit. This corresponds to sending the roots $x_1$ and $y_1$ to $-\infty$ in either case.

We would now like to identify the roots $\{x_1,x_2,x_3,x_4\}$ with the possible choices $\{-1,+1,x_-,x_+\}$, and similarly identify $\{y_1,y_2,y_3,y_4\}$ with the possible choices $\{a,b,y_-,\break y_+\}$. Bearing in mind the ranges in (\ref{box_region}), we have the following two possibilities to obtain the correct signature:
\begin{itemize}
 \item For the case $\brac{a^2-1}\brac{1-b^2}<0$, $\ell^2<0$, the only possibility is to identify $(x_1=x_-,\break x_2=x_+,x_3=-1,x_4=+1)$ and $(y_1=y_-,y_2=a,y_3=b,y_4=y_+)$. These eight roots are then ordered as
  \begin{align}
   y_-<x_-<x_+< a < b <-1<+1< y_+\,. \label{NRBorder_bq}
  \end{align}
 \item For the case $\brac{a^2-1}\brac{1-b^2}>0$, $\ell^2>0$, the only possibility is to again identify $(x_1=x_-,x_2=x_+,x_3=-1,x_4=+1)$ and $(y_1=y_-,y_2=a,y_3=b,y_4=y_+)$. These eight roots are then ordered as
  \begin{align}
   x_-<y_-<a<x_+< -1<b< y_+<+1\,. \label{NRBorder_aq}
  \end{align}
\end{itemize}
These two possibilities respectively describe the dS and AdS cases. The case with no cosmological constant can then be obtained from either case by taking the limit $\ell^2\rightarrow\pm\infty$. In this case, it can be shown that the eight roots are ordered as
  \begin{align}
   y_-=x_-<x_+= a < b =-1<+1=y_+\,.
  \end{align}

\begin{figure}
 \begin{center}
  \begin{subfigure}[b]{0.4\textwidth}
   \centering
   \includegraphics[height=2.4in,width=2.4in]{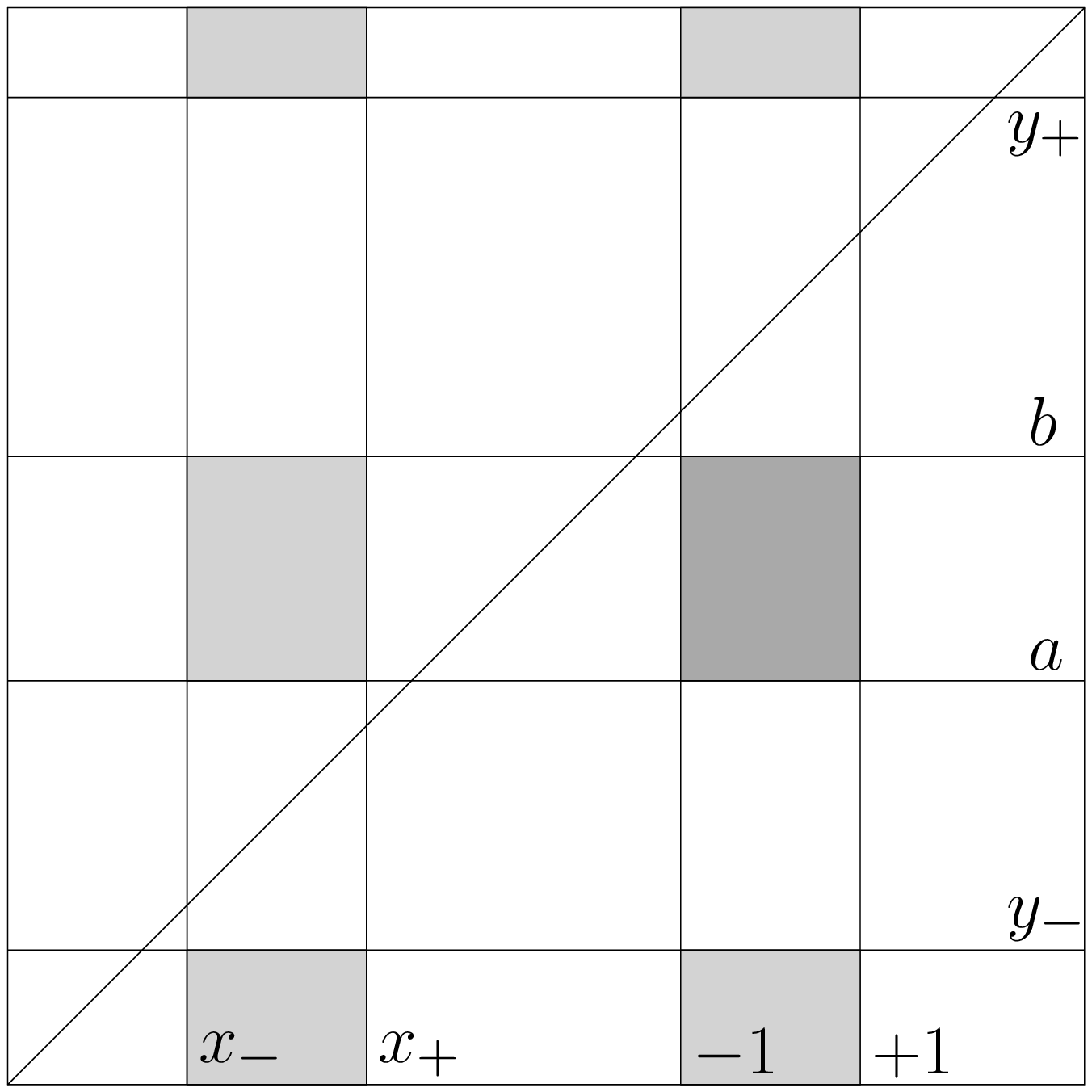}
   \caption{$\ell^2<0$ (dS)}
  \end{subfigure}
~~~  \begin{subfigure}[b]{0.4\textwidth}
   \centering
   \includegraphics[height=2.4in,width=2.4in]{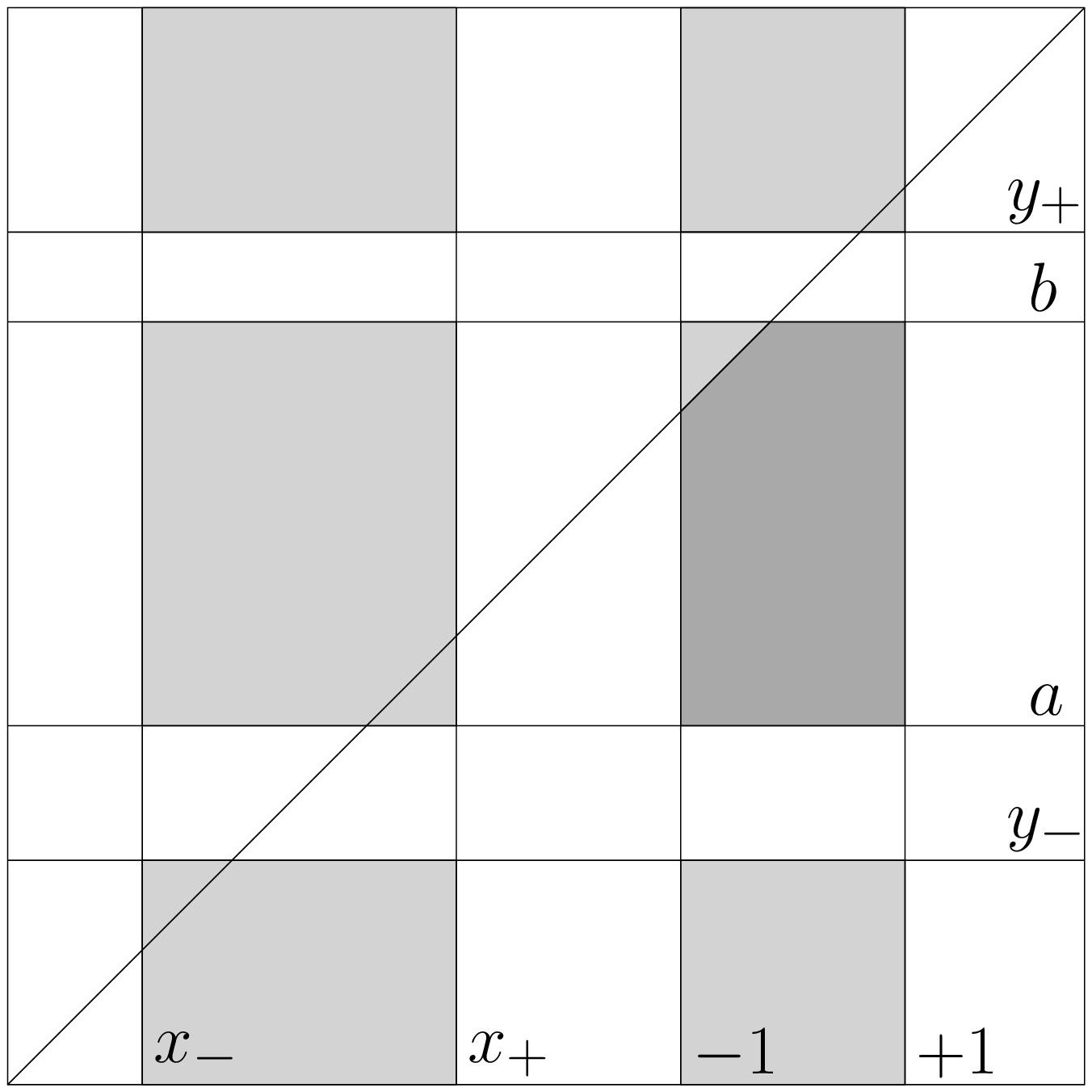}
   \caption{$\ell^2>0$ (AdS)}
   \end{subfigure}\vspace{6pt}
  \begin{subfigure}[b]{0.4\textwidth}
   \centering
   \includegraphics[height=2.4in,width=2.4in]{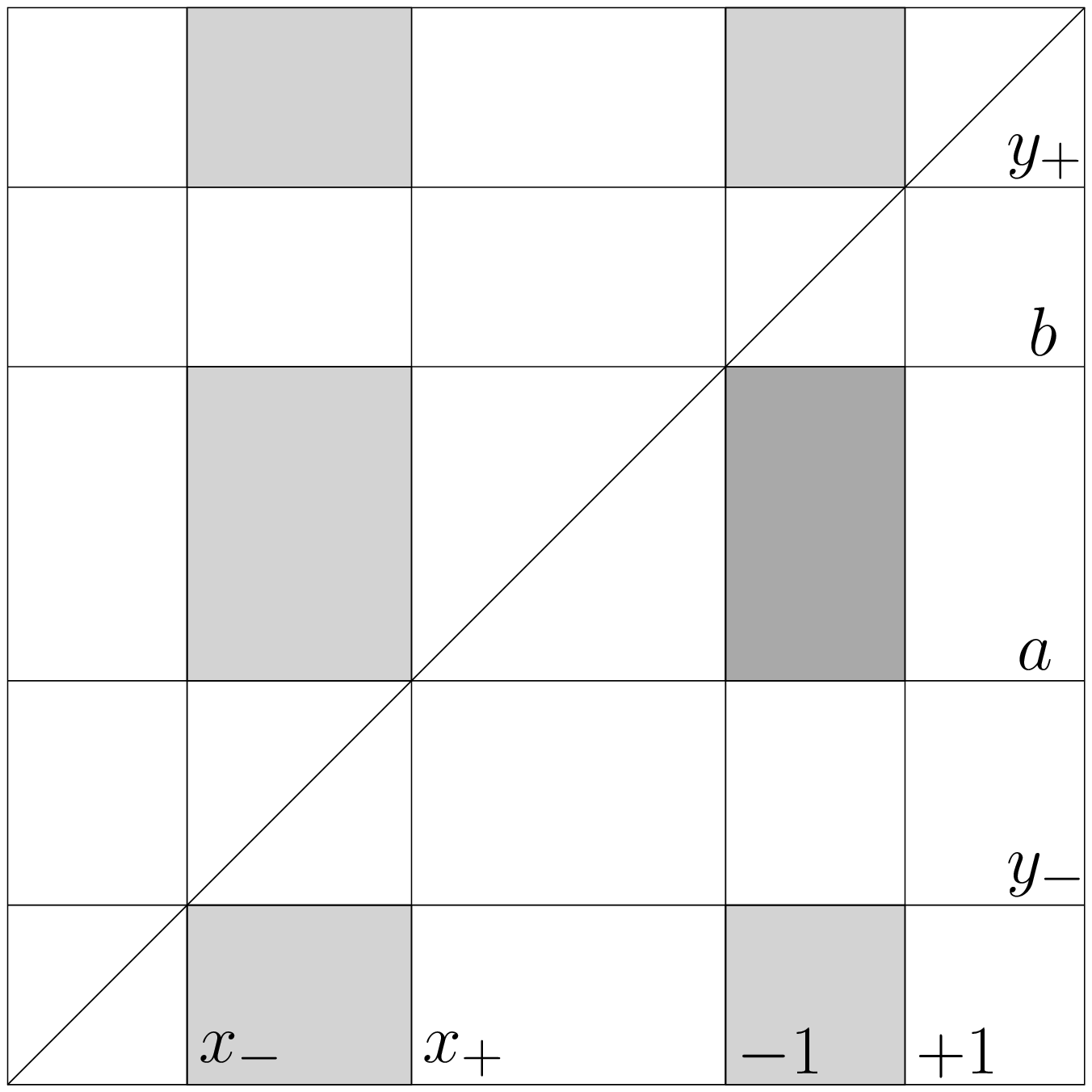}
   \caption{$\ell^2\rightarrow\pm\infty$ (Ricci-flat)}
  \end{subfigure}
 \end{center}
 \caption{The domains of the charged C-metric for the cases (a) $\ell^2<0$ (dS), (b) $\ell^2>0$ (AdS), and (c) $\ell^2\rightarrow\pm\infty$ (Ricci-flat). In each diagram, the $x$-axis is in the horizontal direction, while the $y$-axis is in the vertical direction. Apart from the presence of the extra roots $x_-$ and $y_-$, these diagrams are similar to those in Fig.~\ref{NRBfig2}. The boxes we are interested in are in the darker shade.}
 \label{RBfig2}
\end{figure}

Having determined the order of the roots in the three cases, we can plot them in an $x$-$y$ plot. The domain of interest will then be one of the boxes enclosed by these roots. This is illustrated for each of the three cases in Fig.~\ref{RBfig2} in the darker shade. Apart from the presence of the extra roots $x_-$ and $y_-$, these diagrams are similar to those in Fig.~\ref{NRBfig2}. We remark that asymptotic infinity is again given by the $x=y$ line, and that there are curvature singularities located at the edges of the plot, at $x,y=\pm\infty$.

As in the uncharged limit, the domain of interest may or may not intersect the $x=y$ line, depending on the sign of the cosmological constant. In the dS case, the box will not touch this line at all, while in the Ricci-flat case, one corner of the box will touch this line. In the AdS case, the box will necessarily intersect the line. The part of the box with $y<x$ will then be the domain of interest.

The four edges of the box, $x=\pm1$ and $y=a,b$, can again be interpreted as the axes and horizons of the space-time. Their properties are encoded in the rod structure of the solution, which can be calculated in the standard way. It turns out to be very similar to that of the uncharged case illustrated in Fig.~\ref{rod_diagram}, the only difference being that the surface gravities of the four rods are now given by
\begin{align}
\kappa_{\rm E1}&=a^2+b^2+ab+a+b-1+q^2(a+1)(b+1)\,,\cr
\kappa_2&=\frac{b-a}{2}\left[1+a^2+2ab-q^2(a^2-1)\right],\cr
\kappa_{\rm E3}&=a^2+b^2+ab-a-b-1+q^2(a-1)(b-1)\,,\cr
\kappa_4&=\frac{b-a}{2}\left[1+b^2+2ab-q^2(b^2-1)\right].\label{charged_kappas}
\end{align}
Recall that Rod 2 is a black-hole horizon, while Rod 4 is an acceleration or cosmological horizon. On the other hand, Rods 1 and 3 are axes with possible conical singularities running along them. Now, from the fact that $a+b<0$, we have
\begin{align}
\kappa_{\rm E1}<\kappa_{\rm E3}\,.
\end{align}
This implies that a conical singularity must be present either along Rod 1 or Rod 3, just as in the uncharged case.

We remark that $y=y_-$ can be interpreted as the inner black-hole horizon that is known to be present when the black hole is charged. This is consistent with the fact that the region below the dark-shaded box in each of the plots in Fig.~\ref{RBfig2} represents the region inside the black hole. In the limit of vanishing charge, this inner horizon merges with the black-hole singularity at $y=-\infty$, and the lighter-shaded Lorentzian region between $y=y_-$ and $-\infty$ disappears.

Finally, we recall that in the dS case, it is possible for the roots $x_\pm$ to become complex, in which case they will come together and disappear from Fig.~\ref{RBfig2}(a). The regions enclosed by these two roots, including the lighter-shaded Lorentzian regions, will then also disappear from the diagram. However, this will not affect the domain of interest at all.

\subsection{Parameter ranges}

Let us now briefly study the parameter ranges of the charged C-metric. This will be done separately for each of the three cases of a positive, zero and negative cosmological constant.

\subsubsection*{\underline{Charged dS C-metric}}

Recall that for the uncharged dS C-metric, the temperature of the cosmological horizon is lower than that of the black-hole horizon. It is clear from (\ref{charged_kappas}) that the effect of adding charge decreases the values of $\kappa_2$ and $\kappa_4$. Hence, charging the solution decreases the temperatures of the two horizons; moreover, the temperature of the black-hole horizon decreases faster than that of the cosmological horizon. One might wonder if the temperature of the black-hole horizon can become lower than that of the cosmological horizon. Indeed, this is possible and will occur when $q^2>q_{\rm eq}^2$, where
\begin{align}
q_{\rm eq}^2\equiv1\,.
\end{align}
When $q^2=q_{\rm eq}^2$, the two horizons will have the same temperature, and so are in thermal equilibrium.

As one continues to add charge, the temperature of the black-hole horizon will eventually vanish. This occurs when $q^2=q_{\rm ext}^2$, where
\begin{align}
q_{\rm ext}^2\equiv\frac{1+a^2+2ab}{a^2-1}>q_{\rm eq}^2\,.\label{charged_dS_range1}
\end{align}
It is the maximally charged dS C-metric, with an extremal black-hole horizon. In this case, $y_{-}=a$; in the context of Fig.~\ref{RBfig2}(a), the inner black-hole horizon has merged with the outer horizon. From the order of the roots (\ref{NRBorder_bq}), we see that $x_{\pm}$ must be absent in Fig.~\ref{RBfig2}(a); they are thus complex. 

The full range of parameters for the charged dS C-metric is therefore
\begin{align}
-\infty<a<b<-1\,,\quad 0\leq q^2\leq q_{\rm ext}^2\,.\label{charged_dS_range2}
\end{align}
It can be checked that the temperature of the cosmological horizon remains finite and non-zero in this range. 

\subsubsection*{\underline{Charged C-metric}}

The charged C-metric with no cosmological constant can be obtained from (\ref{Cmetric_charged}) and (\ref{Cmetric_charged_max}) by taking the limit $\ell^2\rightarrow\pm\infty$ and $b\rightarrow -1$, while keeping $H^2$ finite. In particular, it can be obtained from the charged dS C-metric considered above, by taking the lower sign in the limit for $\ell^2$. The parameter range in this case can then be obtained from \Eqref{charged_dS_range1} and \Eqref{charged_dS_range2} in the appropriate limit. It is given by
\begin{align}
-\infty<a<-1\,,\quad 0\leq q^2\leq q_{\rm ext}^2\,,
\end{align}
where
\begin{align}
q_{\rm ext}^2\equiv\frac{a-1}{a+1}\,,
\end{align}
in this case. Like the dS case, the black-hole horizon will be in thermal equilibrium with the acceleration horizon when $q^2=1$, and will become extremal when $q^2=q_{\rm ext}^2$.

As the general form of the solution (\ref{Cmetric_charged}) and (\ref{Cmetric_charged_max}) is no longer valid in this limit, one needs to use an alternative form in this case. For example, it is possible to transform it to the factorised form of the charged C-metric proposed in \cite{Hong:2003gx}, using a coordinate transformation similar to the one described in Sec.~\ref{sec4.2}.

\subsubsection*{\underline{Charged AdS C-metric}}

For the uncharged AdS C-metric, the temperature of the acceleration horizon is lower than that of the black-hole horizon. From (\ref{charged_kappas}), the effect of adding charge decreases the value of $\kappa_2$ but increases that of $\kappa_4$. Hence, charging the solution decreases the temperature of the black-hole horizon but increases the temperature of the acceleration horizon. When $q^2=q_{\rm eq}^2$, the two horizons have the same temperature; when $q^2>q_{\rm eq}^2$, the temperature of the black-hole horizon is lower than that of the acceleration horizon.

The temperature of the black-hole horizon will vanish when $q^2=q_{\rm ext}^2$, where $q_{\rm ext}^2$ is defined as in (\ref{charged_dS_range1}). However, unlike the dS case, there is now no definite order between $q_{\rm ext}^2$ and $q_{\rm eq}^2$. Moreover, there is now the possibility that the temperature of the acceleration horizon can also vanish. This will occur when $q^2=q_{\rm ext2}^2$, where
\begin{align}
q_{\rm ext2}^2\equiv\frac{1+b^2+2ab}{b^2-1}\,.
\end{align}
There is no definite order between $q_{\rm ext2}^2$ and $q_{\rm eq}^2$, or between $q_{\rm ext2}^2$ and $q_{\rm ext}^2$.

To complicate matters, it is also possible for $\kappa_{\rm E1}$ to vanish in the AdS case. This stems from the fact that adding charge to the solution decreases the value of $\kappa_{\rm E1}$. However, recall that even in the uncharged case, it is possible for $\kappa_{\rm E1}$ to vanish for a sufficiently large value of $b$. In that case, the vanishing of $\kappa_{\rm E1}$ can be interpreted as the outer axis becoming a new spatial infinity of the space-time. A similar interpretation can be made for the charged case.

Hence, the full range of parameters for the charged AdS C-metric can be obtained by requiring the conditions
\begin{align}
\label{range_charge_AdS_C1}
-\infty<a<-1<b<+1\,,
\end{align}
and
\begin{align}
\label{range_charge_AdS_C2}
\kappa_{\rm E1}\geq 0\,,\quad \kappa_2\geq 0\,,\quad\kappa_4\geq 0\,,
\end{align}
to hold simultaneously. Note that $\kappa_{\rm E3}$ is always positive if these conditions are satisfied. When there is no charge, this parameter range will reduce to the shaded region in Fig.~\ref{NRBfig3}.

The full parameter range (\ref{range_charge_AdS_C1}) and (\ref{range_charge_AdS_C2}) can be represented by a three-dimensional $a$-$b$-$q^2$ plot, but it is obviously complicated. Thus we will not attempt a full analysis of it here, and be content with just a few general observations. For fixed $a$, the parameters $b$ and $q^2$ are bounded from above by the three physical situations: the outer axis becoming a new spatial infinity when $\kappa_{\rm E1}=0$, the black hole becoming extremal when $\kappa_2=0$, and the acceleration horizon becoming extremal when $\kappa_4=0$. A ``triple point'' when all these conditions are satisfied simultaneously occurs when
\begin{align}
a=-\sqrt{2}-1\,,\quad b=\sqrt{2}-1\,,\quad q^2=1\,.
\end{align}
This special situation describes an extremal black droplet in thermal equilibrium. We mention that black droplets in thermal equilibrium at a non-vanishing temperature have been considered in \cite{Caldarelli:2011wa}.

\section{Classification of all possible domains}
\label{sec6}

We have so far in this paper focussed on space-times described by box-like domains in the $x$-$y$ space. The four edges of the box are just the endpoints of the coordinate ranges in (\ref{box_region}), and describe either axes or horizons in the space-time. Examples of such domains can be found in Fig.~\ref{NRBfig2}(a) and (c), for the dS and Ricci-flat C-metric respectively. For the AdS C-metric, it turns out that the domain is also bounded by the $x=y$ line representing asymptotic infinity; as can be seen from Fig.~\ref{NRBfig2}(b), it still has the general shape of a box, but with one corner cut off.

We have pointed out that the triangular-shaped corner that is cut off by $x=y$ in Fig.~\ref{NRBfig2}(b) represents a new space-time with the correct Lorentzian signature and without curvature singularities. Unlike the AdS C-metric, this new space-time has only one axis ($x=-1$) and one horizon ($y=b$), both extending to infinity. Moreover, the reader would have noticed two other triangular domains present in Fig.~\ref{NRBfig2}(b), one bounded by the lines $x=+1$, $y=c$ and $x=y$, and another bounded by $x=\gamma$, $y=a$ and $x=y$. It turns out that these triangular domains represent black holes with infinitely extended horizons, generalising the hyperbolic and planar black holes of \cite{Mann:1996gj,Mann:1997iz} that are known to exist in AdS space. 

Unlike the box domains, the existence of the triangular domains only requires the existence of one real root each for $P(x)$ and $Q(y)$. The other two roots of $P(x)$ or $Q(y)$ are allowed to become complex. For instance, the third triangular domain mentioned above will continue to exist even if the two roots $y=b,c$ were to come together and disappear from the diagram. Thus the form of the metric (\ref{Cmetric}) used in this paper, whereby two roots of $P(x)$ have been set to be at $x=\pm1$, is not general enough to describe all possible triangular domains. A more natural choice would be to set the one real root of $P(x)$ to be at $x=-1$ (say), and the one real root of $Q(y)$ to be at $y=0$ (say). This new form of the metric will be presented in \cite{CLT}, and will be used to carry out a study of the AdS space-times described by these triangular domains.

At this stage, one might wonder about the existence of other types of domains that describe space-times with the correct signature and without curvature singularities. An inspection of Fig.~\ref{NRBfig2} shows that there are no others, at least when $P(x)$ and $Q(y)$ have three real roots each. The situation in which two roots of $P(x)$ or $Q(y)$ become complex can still be visualised using Fig.~\ref{NRBfig2}, by letting the two roots come together and disappear from the diagram. In this case, two new domains arise; as it turns out, both are from the AdS case Fig.~\ref{NRBfig2}(b).

\begin{figure}[t]
 \begin{center}
  \begin{subfigure}[b]{0.4\textwidth}
   \centering
   \includegraphics[height=2.4in,width=2.4in]{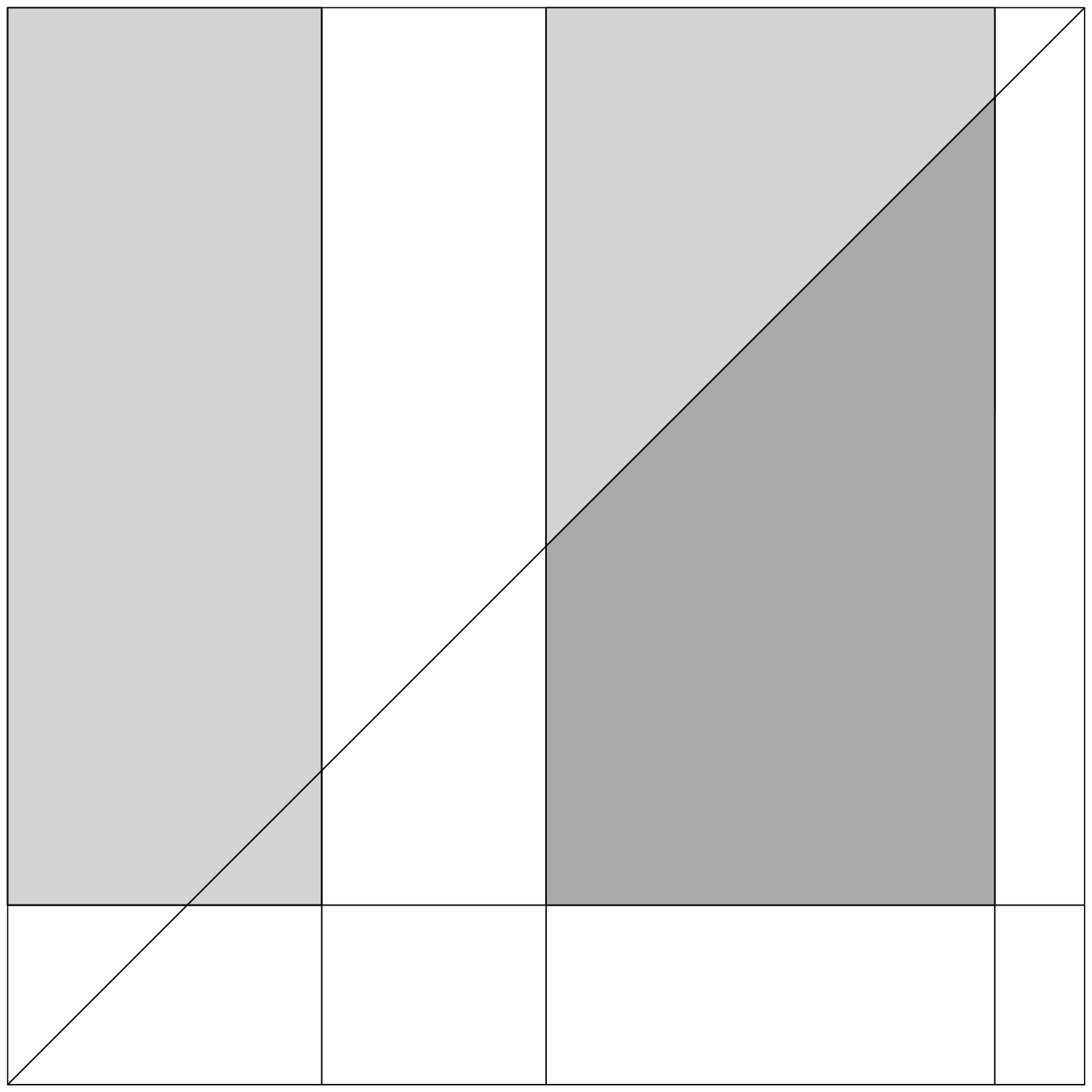}
   \caption{}
   \end{subfigure}\vspace{6pt}
~~~  \begin{subfigure}[b]{0.4\textwidth}
   \centering
   \includegraphics[height=2.4in,width=2.4in]{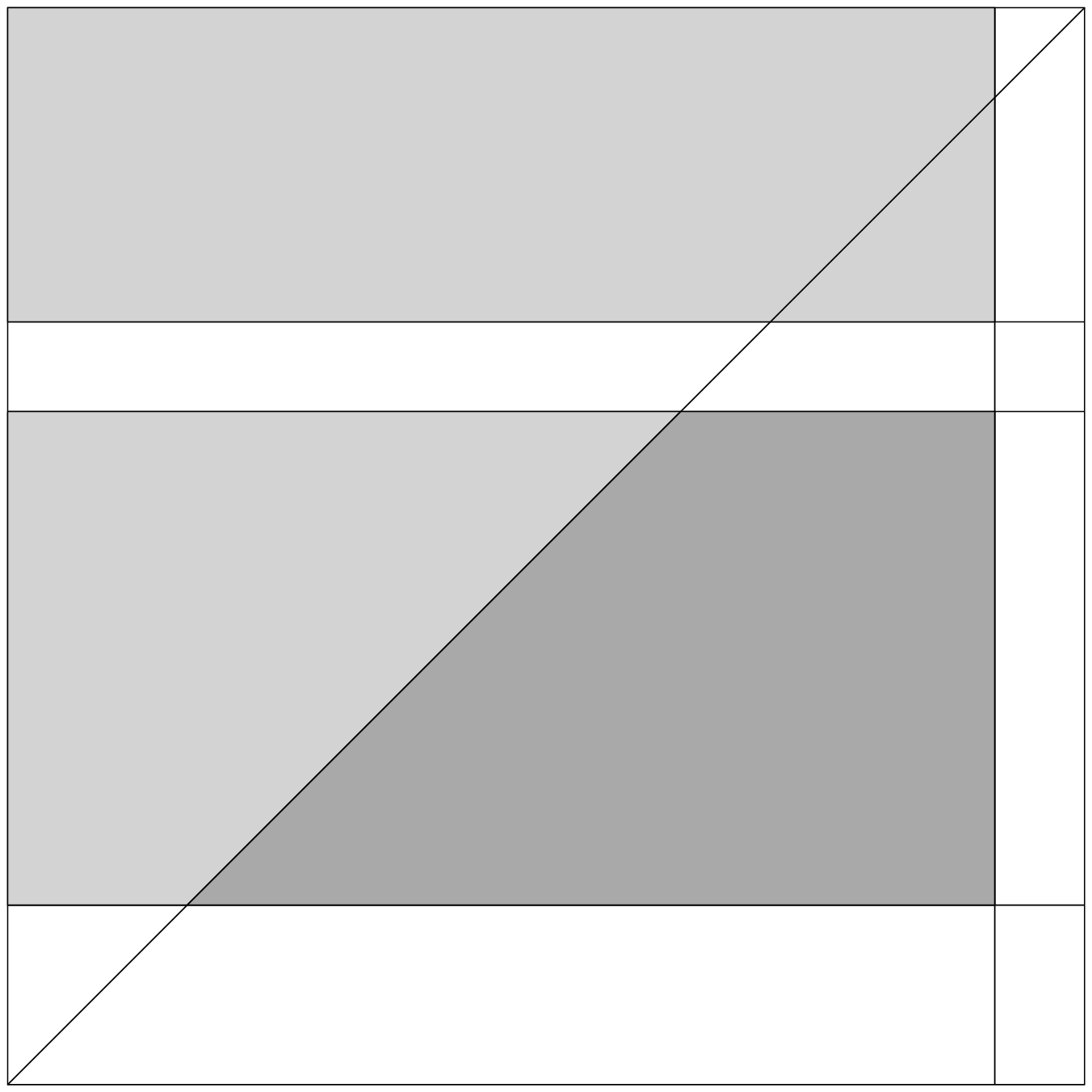}
   \caption{}
   \end{subfigure}
 \end{center}
 \caption{New trapezoidal domains (shown in the darker shade) in the case when (a) $Q(y)$ has one real root and $P(x)$ has three real roots; (b) $P(x)$ has one real root and $Q(y)$ has three real roots. Both these cases are only possible when $\ell^2>0$, i.e., in the AdS case. Note that the two cases (a) and (b) are related by the coordinate symmetry (\ref{coord}).}
 \label{trap_fig}
\end{figure}

The first case occurs when $Q(y)$ has exactly one real root, while $P(x)$ has three real roots. In Fig.~\ref{NRBfig2}(b), this corresponds to letting the roots $y=b,c$ come together and disappear from the diagram. In this case, the box region will merge with the triangular region bounded by $x=+1$, $y=c$ and $x=y$, to form a trapezoidal region as shown in Fig.~\ref{trap_fig}(a). It describes a space-time containing a black-hole horizon and two asymptotic axes; in particular, there is no acceleration horizon present. This solution was analysed in \cite{Podolsky:2002nk}, and was interpreted as a {\it single\/} accelerating black hole in AdS space. In the traditional form of the AdS C-metric (\ref{DL}) below, it corresponds to the case $\tilde A<1/\ell$. We remark that the Schwarzschild-AdS black hole can be recovered as a limit within this trapezoidal domain, when its left and right edges become symmetric.

The second case occurs when $P(x)$ has exactly one real root, while $Q(y)$ has three real roots. In Fig.~\ref{NRBfig2}(b), this corresponds to letting the roots $x=\gamma,-1$ come together and disappear from the diagram. In this case, the box region will merge with the triangular region bounded by $x=\gamma$, $y=a$ and $x=y$, to form a trapezoidal domain as shown in Fig.~\ref{trap_fig}(b). It describes a space-time containing two horizons extending to asymptotic infinity, and one inner axis. This solution has been interpreted by Hubeny et al.\ \cite{Hubeny:2009kz} as a black droplet suspended above a black funnel.

We remind the reader that, like the box domains, the above triangular and trapezoidal domains only represent the static regions of the full space-time. It is possible to extend these regions past the horizon(s) using suitable non-static coordinates.

Finally, we briefly consider the charged case, which turns out to be very similar to the uncharged case. Again, we see that triangular domains are possible only in the AdS case, with four such domains apparent in Fig.~\ref{RBfig2}(b) when all the roots are real. Some of these triangular domains will still exist even when the roots are allowed to become complex. New trapezoidal domains can also arise in this case, generalising those found in the uncharged case.

\section{Discussion}

In this paper, we have proposed a new form of the C-metric with cosmological constant, with the two structure functions partially factorised. Furthermore, this new form is parameterised by roots of the structure functions. As we have seen, this makes the analysis of the solution more forthcoming and transparent than in the traditional form. In particular, it leads to a powerful visual representation of the solution in terms of its so-called domain structure. For the solutions considered in this paper, the domain has the shape of a box; by shifting the upper and lower edges of the box, the different cases of the C-metric with a positive, negative or zero cosmological constant can be obtained.

Besides the box-shaped domains that were the main focus of this paper, there exist domains with other shapes that might describe physically interesting space-times. In Sec.~\ref{sec6}, we classified all the other possible shapes that can emerge from the C-metric with cosmological constant. The two new possible shapes are triangles and trapezoids, and both only occur in the AdS case. It would be interesting to study the physics of the space-times described by these domains in more detail. In a subsequent publication \cite{CLT}, we will show that the triangular domains describe a new class of generalised hyperbolic/planar black holes in AdS space.

A natural extension of the results of this paper would be to include rotation and NUT charge. Such a general C-metric, including electric and magnetic charges, rotation, NUT charge, as well as a cosmological constant, can be identified from the Pleba\'nski--Demia\'nski solution \cite{Plebanski:1976gy}. A form of this solution was proposed by Griffiths and Podolsk\'y in \cite{Griffiths:2006tk,Griffiths:2009dfa}, building upon the results of \cite{Hong:2004dm} for the case of zero NUT charge. Recall that the Pleba\'nski--Demia\'nski solution contains two structure functions, usually denoted as $P(x)$ and $Q(y)$, both of which are quartic polynomials. They satisfy the relation
\begin{align}
Q(x)=P(x)-\frac{1}{\ell^2}(1+x^4)\,,
\end{align}
so there is in general no simple relation between the roots of the two structure functions. In the Griffiths--Podolsk\'y form of the solution, $P(x)$ is partially factorised; on the other hand, $Q(y)$ remains unfactorised. While this form is well suited for identifying the various known limits of the solution, it has the disadvantage that the roots of $Q(y)$ are cumbersome to write down. 

One might wonder if the new form of the C-metric presented in this paper can be extended to this general case. Indeed, we have been able to find a form of the Pleba\'nski--Demia\'nski solution in which both its structure functions are partially factorised. In this form, two roots of $P(x)$ and two roots of $Q(y)$ are regarded as fundamental parameters. Unlike the static case however, there is no longer the freedom to set the two roots of $P(x)$ to specific values. It turns out that the positions of these two roots will determine the amount of rotation and NUT charge that the space-time possesses. For example, if the two roots are equal and opposite in value, then the rotation is non-zero but the NUT charge is zero. All this translates to a compelling picture in terms of the domain structure of the solution, in which the size and location of the box will determine the amount of rotation and NUT charge present. We intend to report on these results in the future.

\section*{Acknowledgement}

This work was partially supported by the Academic Research Fund (WBS No.: R-144-000-333-112) from the National University of Singapore.

\appendix

\section{Equivalence of the new metric to the traditional form}

The C-metric with cosmological constant is traditionally written in the form (see, e.g., \cite{Dias:2002mi,Dias:2003xp,Krtous:2003tc,Podolsky:2003gm,Krtous:2005ej}):
\begin{align}
 \dif s^2&=\frac{1}{\tilde A^2(\tilde{x}-\tilde{y})^2}\brac{{F(\tilde y)}\,\dif\tilde{t}^2-\frac{\dif\tilde{y}^2}{{F(\tilde y)}}+\frac{\dif\tilde{x}^2}{{G(\tilde x)}}+{G(\tilde x)}\,\dif\tilde{\phi}^2},\nonumber\\
  {G(\tilde x)}&=1-\tilde{x}^2-2\tilde m\tilde A\tilde{x}^3-\tilde{q}^2\tilde A^2\tilde{x}^4,\nonumber\\
  {F(\tilde y)}&=\brac{1-\frac{1}{\ell^2\tilde A^2}}-\tilde{y}^2-2\tilde m\tilde A\tilde{y}^3-\tilde{q}^2\tilde A^2\tilde{y}^4. \label{DL}
\end{align}
We would like to see how this form of the metric relates to the one introduced in this paper. For simplicity, we focus only on the uncharged case $\tilde q=0$ here.
To transform (\ref{Cmetric}) into the uncharged limit of (\ref{DL}), we follow \cite{Hong:2003gx} by considering the affine coordinate transformation:
\begin{align}
  x=Bc_0\tilde{x}+c_1\,,\quad y=Bc_0\tilde{y}+c_1\,,\quad t=\frac{c_0}{B}\,\tilde{t}\,,\quad\phi=\frac{c_0}{B}\,\tilde{\phi}\,.
\end{align}
To preserve the form of the metric, we require that
 \begin{align}
  B^2=H^2\tilde A^2,\quad{P(x)}=B^2{G(\tilde x)}\,,\quad{Q(y)}=B^2{F(\tilde y)}\,.\label{BHA}
 \end{align}
Equating the coefficients of the structure functions, we get
\begin{align}
  2\tilde m\tilde A&=-(a+b)Bc_0^3\,,\label{x3}\\
  1&=c_0^2\sbrac{-3(a+b)c_1+a^2+b^2+ab-1},\label{x2}\\
  0 &=3(a+b)c_1^2-2(a^2+b^2+ab-1)c_1 -a-b\,,\label{x1}\\
  1 &=-\frac{1}{B^2}\brac{c_1^2-1}\brac{-(a+b)c_1+a^2+b^2+ab-1},\label{x0G}\\
  1-\frac{1}{\ell^2\tilde A^2}&=\frac{1}{B^2}\brac{c_1-a}\brac{c_1-b}\brac{(a+b)c_1+ab+1}.\label{x0F}
\end{align}
Note that (\ref{x0F}) is not an independent equation, but can be obtained from (\ref{x0G}) and the first equation of (\ref{BHA}).

Now, (\ref{x1}) is a quadratic equation in $c_1$ with the two solutions:
\begin{align}
  -3(a+b)c_1=-a^2-b^2-ab+1\pm\sqrt{(a^2+b^2+ab-1)^2+3(a+b)^2}\,.\label{Bc0c1}
\end{align}
We choose the upper sign, so that (\ref{x2}) has real solutions for $c_0$:
\begin{align}
  \frac{1}{c_0^2}=\sqrt{(a^2+b^2+ab-1)^2+3(a+b)^2}\,. \label{c0soln}
\end{align}
On the other hand, (\ref{x1}) and (\ref{x0G}) can be combined to obtain
\begin{align}
  {B^2}=-\frac{a+b}{2c_1}(c_1^2-1)^2,\label{B2}
\end{align}
which upon substituting (\ref{Bc0c1}) will give an expression for $B^2$ in terms of $a$ and $b$. The right-hand side of (\ref{B2}) is clearly positive if we take the positive sign in (\ref{Bc0c1}), so we are guaranteed real solutions for $B$. Finally,  (\ref{x3}) and the first equation of (\ref{BHA}) can be solved to give expressions for $\tilde m$ and $\tilde A$ in terms of $a$, $b$ and $\ell$. The signs of $c_0$ and $B$ in (\ref{c0soln}) and (\ref{B2}) respectively, should be chosen so as to ensure $\tilde m\tilde A$ is positive by (\ref{x3}).

Turning to the last equation (\ref{x0F}), note that it can be rewritten as
\begin{align}
  1-\frac{1}{\ell^2\tilde A^2}&=\frac{a+b}{B^2}\brac{c_1-a}\brac{c_1-b}\brac{c_1-c},
\end{align}
where $c$ is the third root of $Q(y)$ as defined in (\ref{NRB_roots}). Also note that $c_1$ can be rewritten as
\begin{align}
  c_1=\frac{1}{3}\brac{a+b+c+\sqrt{(a+b+c)^2+3}}.
\end{align}
Now it can be shown that
\begin{align}
  b\leq c_1\leq c\,,
\end{align}
with the equalities holding if and only if $b=c$. From the ordering of the roots $a<b<c$ and the condition (\ref{aplusb}), we conclude that
\begin{align}
  1-\frac{1}{\ell^2\tilde A^2}\geq 0\,.\label{cond1}
\end{align}
This is precisely the condition identified in \cite{Dias:2002mi} for ${G(\tilde x)}$ and ${F(\tilde y)}$ to have three real roots each. The case of equality in (\ref{cond1}) corresponds to the situation in which two roots of ${F(\tilde y)}$ become degenerate. Recall that this is only possible in the AdS case, and it corresponds to the upper bound $b=-a-\sqrt{a^2-1}$ in (\ref{AdS_range1}); it is in fact the solution considered in \cite{Emparan:1999wa}.

\bigskip\bigskip

{\renewcommand{\Large}{\normalsize}
}

\end{document}